\documentclass[english]{article}
\usepackage[T1]{fontenc}
\usepackage[latin9]{inputenc}
\usepackage{amstext}
\usepackage{amssymb}
\usepackage{graphicx}
\usepackage{esint}

\makeatletter

\providecommand{\tabularnewline}{\\}

\newcommand{\lyxaddress}[1]{
\par {\raggedright #1
\vspace{1.4em}
\noindent\par}
}

\makeatother

\usepackage{babel}
\begin{document}

\title{\textbf{\Large{}The role of the Havriliak-Negami relaxation in the
description of local structure of Kohlrausch's function in the frequency
domain. Part I}}

\author{{\normalsize{}J.S. Medina,}%
\thanks{{\footnotesize{}tlazcala@yahoo.es}%
}{\normalsize{} R. Prosmiti,$^{1}$ and J.V. Alemán$^{2}$}}

\maketitle

\lyxaddress{\emph{\footnotesize{}$^{1}$Instituto de Física Fundamental, IFF-CSIC,
Serrano 123, Madrid ES-28006, Spain}}

\lyxaddress{\emph{\footnotesize{}$^{2}$Departamento de Química, Facultad de
Ciencias del Mar, ULPGC, Campus Universitario de Tafira, Las Palmas
de G. Canaria ES-35017, Spain}}
\begin{abstract}
An improved approximation via Havriliak-Negami (HN) functions to the
Fourier Transform (FT) of certain Weibull distributions, $-\psi_{\beta}$,
(the time derivative of the Kohlrausch-Williams-Watts function), is
given for a large interval of frequencies: $\omega/2\pi\in[0,10^{12}]$
if $0<\beta\leq1$ and $\omega/2\pi\in[0,10^{7}]$ if $1<\beta\leq2$.
The model is free from the usual numerical distortions, or restrictions
associated to sampling step and finite size, present in similar adjustments
to complex relaxation functions. Further indicates that the identification
of (FT) Weibull data with a double HN approximant, $\psi_{\beta}\simeq\mathcal{A}p_{2}HN$,
is quite exact locally even though the parameters involved should
vary \emph{adiabatically} with the frequency, \emph{i.e. }$\{\alpha_{1,2},\gamma_{1,2},\tau_{1,2},\lambda\}(\omega)$.
This fact is the base for the high sensibility of the parameter models,
as functions of $\beta$, to the available data and sampling associated
errors. Presumably is also the root of the different proposals for
asymptotic traits of $\alpha\cdot\gamma(\beta)_{1,2}|_{\omega\rightarrow\infty}$
functions found in scientific literature.
\end{abstract}

\section*{Introduction}

The Kohlrausch relaxation function, $\phi_{K,\beta}(t)\equiv\exp-t^{\beta}$,
$0\leq t<\infty$, $\beta\leq1$,%
\footnote{As the rescaling is always possible we will use here dimensionless
variables solely, for both times and frequencies, \emph{i.e. $t/\tau_{K}\mapsto t$}
and\emph{ $\omega\tau_{K}\mapsto\omega$.}%
} initially proposed to explain the discharge of capacitors \cite{Kohl 1854},
has found later several applications in many fields not directly related
to electrostatics. And while it was half-forgotten in the scientific
literature which was more worried about practical applications, it
did not dissapear at any time of the mathematical heritage due to
the deep connections that such function and its sequel, $\beta>1$,
have with number theory \cite{Hard 1920,Burw 1924,Bakh 1933,Mari 1959,Seno 1996,Kami 1998},
stable distributions in probability \cite{Boch_1 1937,Boch_2 1937,Wint 1941,Humb 1945,Poll 1946,Berg 1952,Zolo 1986},
theta function extensions \cite{Wint 1947,Wint 1956}, or the special
functions obtained from certain kernels of Mellin-Barnes type transformations
employed in fractional calculus \cite{Fox 1928,Gloe 1993,Cons 1997,Metz 1998,Hilf 2002,Chen 2004,Pari 2010}.
However it is a result of its use by Williams and Watts \cite{Will 1970},
in the field of characterization of dielectric properties of polymers
\cite{Will 1970,Will 1971,Lind 1980}, that its employment became
ubiquitous in many areas of Physics and Chemistry. Hereafter not only
associated with specific electrical phenomena but a plethora of them
in the most diverse fields. It emerges in luminiscence \cite{Hage 1987,Boes 1989,Lin 1994,Benn 2001,Berb_1 2005,Berb_2 2005,Kuzn 2008},
supercooled glasses \cite{Phil 1996,Hans 1997,Debe 2001}, capacitors
\cite{Kohl 1854,Macd 1986,Hals 1991,Macd 1997}, rheology \cite{Gurt 2001,Ande 2004,Baeu 2005},
spin glasses \cite{Phil 1996}, torsion of galvanometric threads \cite{Kohl 1863},
dielectric spectroscopy \cite{Boes 1989,Macd 1997,Mont 1984,Boeh 1993,Kaat 1996,Ferg 2006,Vici 2009,Schr 2010},
magnetic resonance \cite{Chun 1991,Weis 1994,Benn 2003}, autocorrelation
functions in molecular dynamics \cite{Hage 1987,Lin 1994,Chun 1991,Medi 2011},
econophysics \cite{Lahe 1998,Wu 2007}, protein folding \cite{Metz 1998,Kuzn 2008,Naka 2004,Fier 2007,Frau 2009}
and even astrophysics \cite{Dobr 2007}.

Since it shows up in many branches of physics of complex systems \cite{Hals 1991,Frau 2009,Macd 1987,Phil 1988,Sche 1991}
and soft matter \cite{Phil 1996,Gurt 2001,Boeh 1993,Vici 2009,Sast 1998,Schr 1998,Kahl 2010},
among others, seems not be linked to a precise interaction or a particular
physico-chemical phenomenon, instead it is more appropriate to associate
it to an emergent property. In such a case any ensemble of interacting
elements organized in multiscale clusters whose local relaxations,
or restructuring bonds, proceed to jumps with random wait times of
type $t^{-a}$ should present an autocorrelation, or decay, of Kohlrausch's
nature \cite{Hans 1997,Gurt 2001,Baeu 2005,Sche 1991,Palm 1984,Shle 1984,Will 1985,Wero 2013}.
Thus its hierarchical structure, and possibly experimental difficulties,
advises us to study this relaxation not only through its response
along time but also in other spaces of representation, such as of
frequency \cite{Macd 1986,Schr 2010,Medi 2011,Kahl 2010,Alva 1991,Alva 1993,Havr 1995,Scha 1996,Diaz 2000}.
An exhaustive account of the distribution of intermolecular configurational
energies, characteristic relaxation times --real or virtual--, and
frequency response will give us a valuable set of techniques to tie
in different analytical and laboratory procedures. And indeed such
possibilities are available since the Transforms of Fourier and Laplace,
($\omega-$space and $\tau-$space respectively), exist for the case
$0<\beta\leq2$ \cite{Boch_1 1937,Boch_2 1937,Wint 1941,Poll 1946,Berg 1952}.
Either as a convergent series or an asymptotic one, according to subcases
$0<\beta\leq1$ and $1<\beta\leq2$ and depending on we are talking
about high or low frequency \cite{Boch_1 1937,Boch_2 1937,Wint 1941,Poll 1946,Berg 1952,Wutt 2009}.

Unfortunately these series present several problems of convergence
mainly near zero, and in particular the best known of them shows up
an essential singularity and the others are simply asymptotic \cite{Wint 1941,Poll 1946,Berg 1952,Will 1971,Lind 1980,Wutt 2009},
\emph{i. e. }non convergent as a whole, and with coefficients associated
to the special function gamma, $\Gamma$, over the entire reals which
makes calculations more tedious than in other simpler expressions
\cite{Mont 1984,Chun 1991,Koiz 1978,Dish 1985}. This makes the sum
of such series a burdensome task which is possible to get round with
numerical methods or resummation of series \cite{Weis 1994,Wutt 2009,Snyd 1999,Helf 1983}.
Nowadays both techniques allow a direct and quick calculation of Laplace
and Fourier Transforms with enough accuracy to be of great utility
in spectral filtering, or analysis, of laboratory data \cite{Wutt 2009,Snyd 1999}.
Nevertheless could be missed a concise mathematical formula to give
account, even approximately, of the mentioned transforms to accelerate
the calculations or evaluate repeatedly such functions \cite{Medi 2011,Kahl 2010,Alva 1991}.
The latter is usually a must in optimization since the multiple evaluation
of objective functions with variational parameters increases drastically
computational loads. Genetic algorithms are a good example of this
problem. On the other hand a short analytical expression makes it
easier to compare with the most common mathematical functions in the
complex domain which are already used in the field of dielectric relaxation,
such as Debye \cite{Deby 1913}, Cole-Cole \cite{Cole 1941}, Cole-Davidson
\cite{Davi 1951}, Havriliak-Negami \cite{Havr 1966,Havr 1967} and
others.

We use the notation $\chi_{\beta}(\omega)=$ $\int_{0}^{\infty}e^{-i\omega t}\phi_{K,\beta}(t)\mathrm{d}t$
for the one-sided Fourier transform of the Kohlrausch function and
$\psi_{\beta}(\omega)=$ $-\int_{0}^{\infty}e^{-i\omega t}\frac{d}{dt}\phi_{K,\beta}(t)\mathrm{d}t$
for minus the transformation of the Weibull distribution. They are
both related in $\omega-$space by the algebraic closure relationship
$\psi_{\beta}(\omega)+i\omega\chi_{\beta}(\omega)$ $=1$. In this
way it is indistinct to work with $\psi_{\beta}$ or $\chi_{\beta}$,
the former is however preferable since it allows to remove a pole
and obliterates any singularity in the neighborhood of $\omega\approx0$
while using the resummation formula for the original series of inverse
powers of $\omega^{\beta}$ which describes $\chi_{\beta}$, (for
more details see Ref. \cite{Weis 1994}). Other advantages of employing
$\psi_{\beta}$ are homogeneity, boundedness and similarity of this
family of functions if the shape parameter $\beta$ is varied, as
we pointed out previously \cite{Medi 2011}. We will improve this
reference, and others similar in nature \cite{Macd 1986,Schr 2010,Alva 1991,Alva 1993},
extending the limits imposed by numerical samplings in $t-$space
and making consistent the description of high frequencies decays for
$\psi_{\beta}$ with a unique set of strict Havriliak-Negami functions
when $\beta\leq1$, and with set of same kind although parametrically
extended if $\beta>1$ \cite{Medi 2011}. Besides we will show how
the apparently supernumerary parameters of the Havriliak-Negami function
\cite{Havr 1966,Havr 1967}, $\frac{1}{(1+(i\omega\tau_{HN})^{\text{\ensuremath{\alpha}}})^{\gamma}}$,
$0<\alpha,\gamma\leq1$, when approximating the mentioned transform
are uniquely determined by the parameter $\beta$. Thus these models
with such strong dependency $\{\alpha,\gamma,\text{\ensuremath{\tau}}\}(\beta)$
are equated with others quite new of less parameters \cite{Kahl 2010}.

\section{Analytical considerations}

\subsection{The asymptotes of the data}

As it was shown \cite{Medi 2011}, a double approximant of Havriliak-Negami
functions describe fairly well the FFT of the derivative of the Kohlrausch
function as well as the Cole-Davidson-Kohlrausch family \cite{Medi 2011,Kahl 2010}.
Being the main source of error an small impairment of the fit in the
low frequencies. Nevertheless as the FFT is obtained from the original
function by means of a finite window sampling, which in the Fourier
space is reflected as a convolution with a sinc function, a round
off of the spectra in the high frequency border of the domain is obtained.
Consequently the asymptotic behaviour of the original function and
the approximant will differ, at least in the high frequency domain
limits. Besides, due to the smearing of the peaks when the numerical
transformation is involved, a difference between the theoretical FT
and that of the numerically sampled function is expected. This is
why the double approximant fitted to the numerical transformation
possibly increments even more than its mathematical limitations the
difference with the FT of $\beta t^{\beta-1}e^{-t^{\beta}}$ at the
low frequencies. The question raised is then how sensitive are the
parameters obtained in the optimization to this deformation, inasmuch
as the asymptotic behaviour of the Kohlrausch's (or Weibull's \cite{Weib 1951,Rinn 2009})
Fourier Transform has been indexed to them (\emph{v.gr.} $\alpha_{i}\cdot\gamma_{i}=\alpha_{i}\cdot\gamma_{i}(\beta)$
and the rest are functions of $\beta$ too) \cite{Alva 1991,Alva 1993,Havr 1996}.

To assess the influence of finite-size deformation in the parameters
obtained by fitting a couple of Havriliak-Negami functions to the
Fourier Transform, $\psi_{\beta}(\omega)$, we have performed two
different simulations of it with a general purpose mathematics package
(Mathematica\textsuperscript{TM}). The aim is to obtain accuracy
and avoid numerical oscillations as far as possible, the latter a
characteristic of the integrals involved that makes difficult results
at very high frequencies, as well as to get the proper asymptotic
behaviour of the tails.

As it was for the FFT case the first series comprehends frequencies
from $\nu=0$ to $\nu$ = 500.0005 in steps of $\delta\nu$ = 1/999.999,
being $\omega=2\pi\nu$ \cite{Medi 2011}. The range of parameter
$\beta$ simulated is (0, 2.00{]}, and two subintervals of different
traits are analyzed independently. One $\beta\in$(0,1.00{]} the other
$\beta\in$(1.00,2.00{]} with test points chosen as \{0,1\}.\emph{xy
}with \emph{x} = \{0,1,2,3,4,5,6,7,8,9\}\emph{ }and \emph{y} = \{0,2,5,8\}
and the end point $\beta$ = 2.00.

For the second series the $\beta$ points are the same aforementioned
but the choice for the interval of frequencies is different for each
of the cases $\beta\leq$1 and $\beta>$1. So if $\beta\leq$1, $\nu\in${[}0,10\textsuperscript{12}{]}
and for $\beta>$1, $\nu\in${[}0,10\textsuperscript{7}{]} and the
reason for these distinct intervals is the increasing numerical noise
that overshadows the signal. This makes any calculation useless beyond
$\nu$ = 10\textsuperscript{7} for $\beta>$1 and mostly for $\beta\gtrsim$1.90.
Besides the steps of frequencies are not homogeneous in the whole
interval as they were in the first series; they are now in this way
only in logarithmic scale. This is, in the interval $\nu\in${[}0,10\textsuperscript{-5}{]}
the step takes the value $\delta\nu$ = 10\textsuperscript{-8}, and
in each interval $\nu\in$($10^{a_{i}},10^{a_{i}+1}${]} the increment
is $\delta\nu_{i}=9*10^{a_{i}-3}$ with $a_{i}$ any integer of set
\{-5,-4,...,-1,0,1,...,11,12\} if $\beta\leq$1, or set \{-5,-4,...,-1,0,1,...,6,7\}
if $\beta>$1. So with this second series it is possible to sample
a wider domain not affordable with the fine grain step we used in
the first series, the drawback here is the different implicit 'weight'
the points acquired in both series when the fitting to a double Havriliak-Negami
approximant is made.

The reason for such a large domain of frequencies it is not the reconstruction
purpose, as was shown that a domain of low to medium frequencies,
(\emph{i.e.} $\nu\in${[}0,500.0005{]}), were enough to depict the
function $e^{-t^{\beta}}$ quite accurately by means of inverse FFT
of $\chi_{\beta}(\omega)$ \cite{Medi 2011}. Nor, of course, the
physical reconnaissance by means of dielectric spectroscopy as usually
is done in polymer or metallurgical sciences because just one type
of relaxation is rarely presented in isolation. Usually different
kind of relaxations, associated to several size scales and diverse
phenomena, are registered together in contiguous domains of frequency
spoiling each diagram the head and tail of its neighbors. Rather we
are interested in tails from a mathematical point of view, since a
proper knowledge of the function --and this includes the exact tail
shape--, will allow us to concoct the more precise approximant possible
in the whole domain, and which should include the difficult-by-series
description at low frequencies, \emph{i.e. }near\emph{ }$\nu\sim0$.

The modulus of function $\psi_{\beta}(\omega)$ presents the following
asymptotic behaviour in the domain of frequencies: $|\psi_{\beta}(\omega)|$$\sim$1
when $\omega\rightarrow0$ and $|\psi_{\beta}(\omega)|$$\sim$$\Gamma(\beta+1)/\omega^{\beta}$
as $\omega\rightarrow\infty$ being monotonically decreasing in the
intermediate values. Nevertheless the interval of frequencies in which
a significant amount of this dropping takes place depends strongly
with the value of parameter $\beta$. While for values near to zero,
$\beta\rightarrow0$, the decrement of the signal is slow and mild,
near to one, $\beta\sim1$, is quicker and sharper. For example in
the interval $\nu\in[0.001,500.0005]$, the initial values of $|\psi_{\beta}(\omega)|$
are: (0.665237, 0.793929, 0.999607, 0.999973, 0.999980) and the respective
percentage drops are (14.462\%, 56.382\%, 98.429\%, 99.931\%, 99.968\%),
at the end of such domain, when $\beta$ is set to (0.02, 0.10, 0.50,
0.90, 1.00). This suggests important features about the functions
$|\psi_{\beta}(\omega)|$, namely the plateau near $\nu\sim0$ is
growing in length as $\beta\rightarrow1$, the mentioned evolution
towards a steeper function, and consequently the less important contribution
of the tails to the decrement as $\beta$ varies. Such behaviour is
held and even exacerbated in the interval $1<\beta\leq2$, not only
because the drop is greater --(99.9851\%, 99.9992\%, 99.9999\%) for
$\text{\ensuremath{\beta\in}\{1.10, 1.50, 2.00\}}$, with starting
values (0.999985, 0.999992, 0.999996) of $|\psi_{\beta}(\omega)|$--
but also a sudden change in the decreasing of the functions is perceived.
\begin{figure}[h]
\centering{}\includegraphics[width=0.8\columnwidth]{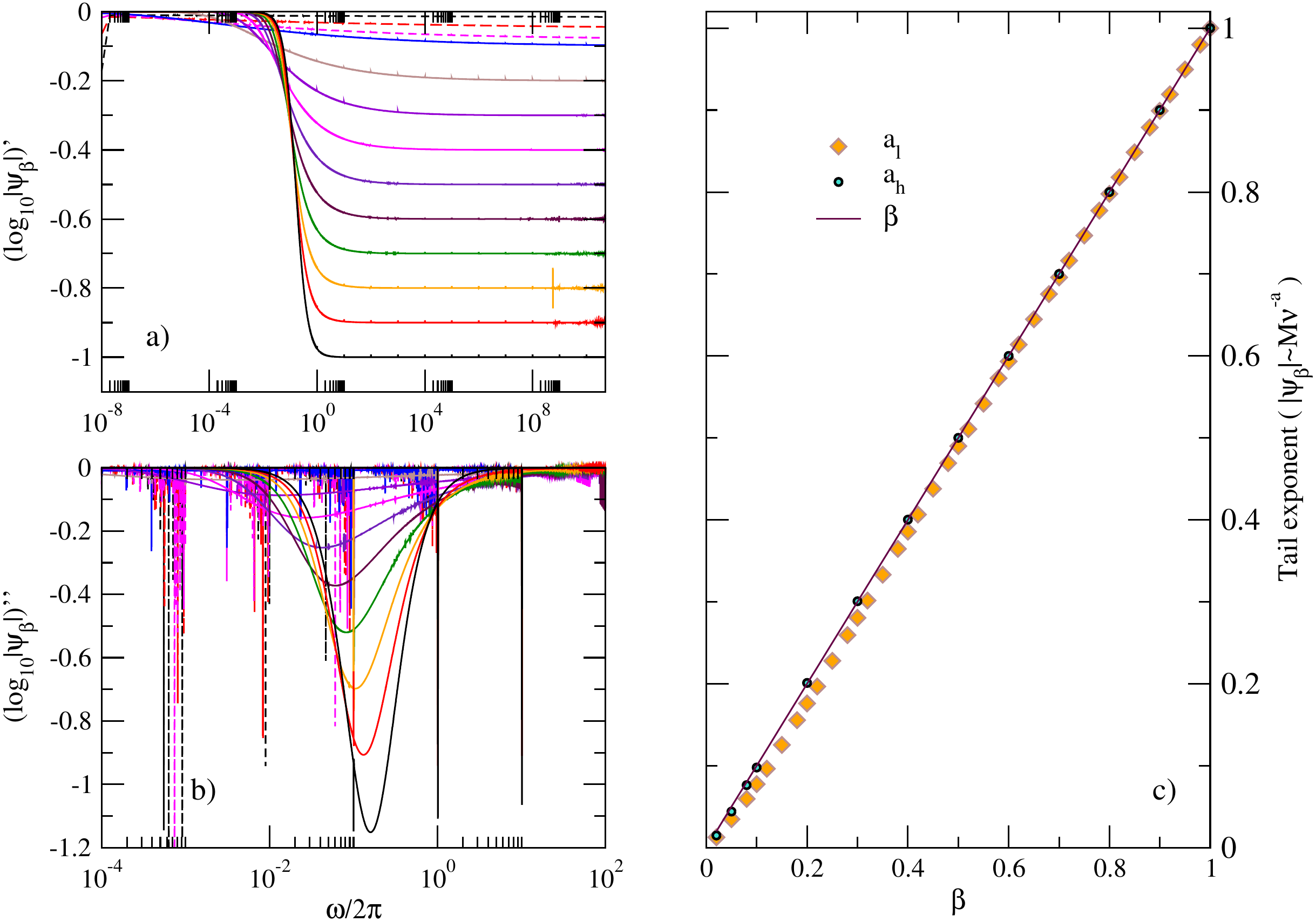}\protect\caption{{\footnotesize{}\label{Figure 01}a)}{\small{} }{\footnotesize{}Logarithmic
derivative of $\log_{10}|\psi_{\beta}(\omega)|$ }\emph{\footnotesize{}vs.
}{\footnotesize{}$\log_{10}(\omega/2\pi)$. b) Logarithmic second
derivative of $\log_{10}|\psi_{\beta}(\omega)|$ }\emph{\footnotesize{}vs.
}{\footnotesize{}$\log_{10}(\omega/2\pi)$. c) Exponent $a$ of fitting
$|\psi_{\beta}(\omega)|\sim\omega^{-a}$ to data tail. Subindex $l$
refers to a sampling of low to medium frequencies and subindex $h$
to a high frequencies one. Solid line would imply $a\equiv\beta$.
See details in text. $\beta\leq1$.}}
\end{figure}

In order to avoid that these characteristics of the $|\psi_{\beta}|$
functions could make seem them belonging to a irregular family of
functions ruled by a parameter, we shall change the scale of the abscissae
axis and we will draw our attention to the first and second derivatives
of such functions in the new scale. The choice for the axis is the
decimal logarithmic scale of variable $\omega/2\pi$ and the first
derivative to look at is $(\log_{10}|\psi_{\beta}|)'$$\equiv\frac{d\log_{10}|\psi_{\beta}|}{d\log_{10}\nu}$.
Obviously $(\log_{10}|\psi_{\beta}|)''$ represents the second derivative
$\frac{d^{2}\log_{10}|\psi_{\beta}|}{d(\log_{10}\nu)^{2}}$.

With this depiction it is clear that each $(\log_{10}|\psi_{\beta}|)'$
looks like a step whose height is greater as $\beta$ approaches 1
from 0, evolving the figure from a slope to a sharp jump around the
same interval of frequencies, $\nu\in(10^{-2},10^{1})$, and being
almost constant outside it. On the other hand the second derivative
$(\log_{10}|\psi_{\beta}|)''$ is a smooth peak, almost symmetrical,
and of contracting half-width, which maximum moves from $\nu_{max}\approx10^{-2}$
to $\nu_{max}\approx2\times10^{-1}$ as $\beta$ goes to value $1.00$
from value $0.30$, (see figure \ref{Figure 01}\emph{a} and \emph{b}).

That is, there is a plateau for each function $|\psi_{\beta}|$, at
low frequencies, and suddenly it bends in a neighborhood of $\nu\sim10^{-1}$,
till it reaches a constant potential decaying tail, which is reflected
in the graphic of the first derivative as a very quick approach to
its horizontal asymptote of value $\beta$. The greater the value
of $\beta$ the quicker the approximation to the horizontal, \emph{i.e.}
for small values of $\beta$ a larger sampling of frequencies is necessary
to fit the function $|\psi_{\beta}|$ with a potential decay $M\nu^{-a}$
and to get $a\approx\beta$. To quantify how long this frequency interval
should be, we have split the domain of modulus of $\psi_{\beta}$
in two regions, one corresponding to the plateau and fall, which we
discarded, and the other corresponding to the tail, being the mark
at $\nu\approx20$ and the end at $\nu=500.0005$ --denoted with subindex
\emph{l }--, or at $\nu=10^{12}$ --denoted \emph{h }--, and we proceeded
to adjust the tail to the generic function $M\nu^{-a}$, with $M$
and $a$ the parameters of the fit. (See figure \ref{Figure 01}).
It is shown that both series of exponents, $a_{l}$ and $a_{h}$,
follow closely the line corresponding to an ideal exponent $\beta$.
The first series, that considering frequencies till $\nu=500.0005$,
shows a slight bulge in the interval 0.10 $\leq\beta\leq$ 0.30 where
the absolute error is greater although the relative error $(\beta-a_{l})/\beta$
is a monotonously decreasing result in the path from $\beta$ = 0.02
to $\beta$ = 1.00. As we pointed out, this is a consequence of the
slower relaxation of the logarithmic slope, $(\log_{10}|\psi_{\beta}|)'$,
towards its asymptote for small values of beta and it doesn't indicate
at all a qualitative change in the behaviour of the modulus of $|\psi_{\beta}|$
that could break the trend of tail decaying as $\nu^{-\beta}$. Nor
it suggests the need of a functional description for the low frequency
part of $|\psi_{\beta}|$ for small betas different than for bigger
ones. The second series corroborates this though as the absolute error
$|\beta-a_{h}|$ is barely perceptible for the values of $\beta$
= 0.02, 0.05, 0.08 or 0.x0 with x $\in\{1,2,...,9\}$ and $\beta=1.00$.
Needless to say that the relative error is even smaller than the previous
case for each $\beta$, (see figure \ref{Figure 01}\emph{c}).

Following this a question raises when an adjustment to a probe function,
(\emph{v. gr.} a sum Havriliak-Negami functions), is tried. Is the
test function 'flexible enough' to describe the whole range of frequencies?
Or is it better to describe the original data with two 'maps', one
for low frequencies and the other for high ones and matching them
conveniently in the frontier?

The question is not trivial, on one hand because the weight of the
low frequency part competes with the weight of the high frequency
part and the former distorts the potential behaviour we have just
described, moving away from $-\beta$ the value of the exponent of
tails. And in the other hand numerical samplings and manipulations,
as in FFT, broke the tail trend and rounded it off affecting consequently
the obtained values of parameters. These issues will be addressed
in the following pages by means of sums of Havriliak-Negami functions
fitted to data calculated almost symbolically.
\begin{figure}
\centering{}\includegraphics[width=0.8\columnwidth]{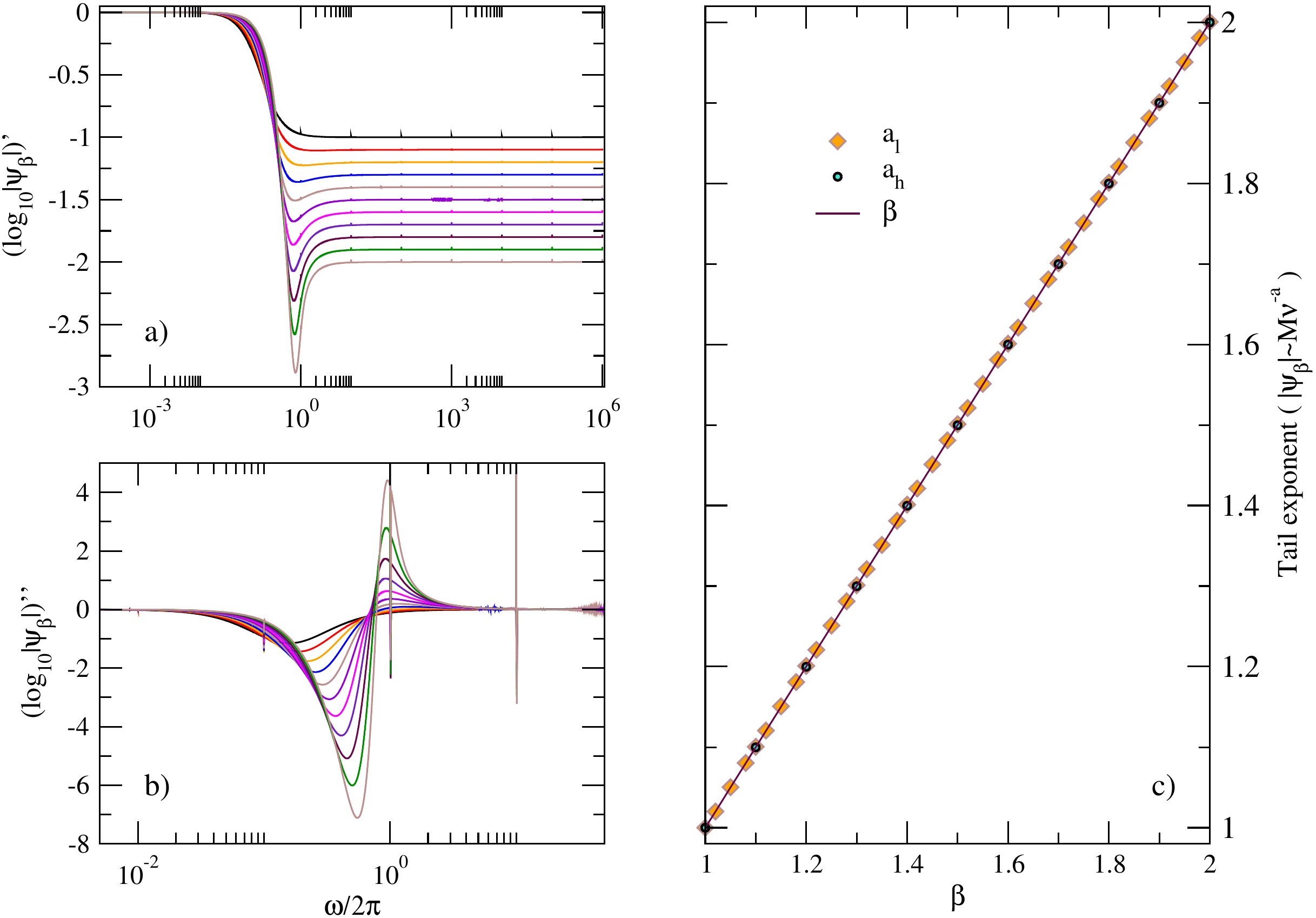}\protect\caption{{\footnotesize{}\label{Figure 02}As in figure \ref{Figure 01}: a)}{\small{}
}{\footnotesize{}Logarithmic derivative of $\log_{10}|\psi_{\beta}(\omega)|$
}\emph{\footnotesize{}vs. }{\footnotesize{}$\log_{10}(\omega/2\pi)$.
b) Logarithmic second derivative of $\log_{10}|\psi_{\beta}(\omega)|$
}\emph{\footnotesize{}vs. }{\footnotesize{}$\log_{10}(\omega/2\pi)$.
c) Exponent $a$ of tail for fitting $|\psi_{\beta}(\omega)|\sim\omega^{-a}$.
Subindexes $l$ and $h$ refer to two different sampling ranges of
medium and high frequencies, respectively. Solid line, again, signalizes
ideal asymptotic tail $a\equiv\beta$. See text for explanations.
$\beta>1$.}}
\end{figure}

However some distinctions should be pointed out with the case $1<\beta\leq2$,
before to continue. Although both are the only ones with a positive
density function of times of relaxation for all the stretched-squeezed
exponentials \cite{Boch_1 1937,Boch_2 1937}, $\exp(-t^{\beta})$,
(\emph{i.e. }with $\beta\in$$(0,\infty$)), they behave differently
in the derivative and both show an obvious parametric discontinuity
in the $t-$space. While for $\beta<1$ the minus derivative, $\beta t^{\beta-1}e^{-t^{\beta}}$,
shows a singularity at $t=0$, the case $\beta>1$ presents a regular
and continuous zero value though it pays the price of losing monotone
decreasing behaviour of the former. Now it displays an increasing
monotone growth and subsequently a decreasing monotone tail which
saves the character of the positivity of the density function \cite{Boch_1 1937,Boch_2 1937}.
They could be cases of a very different appearance or even to show
a break in the continuity after doing the Fourier Transform for both
but the real fact is that the parametric family of the first derivative
transform, $\psi_{\beta}(\omega)$, does not present any oddity. Nonetheless
the instance $\beta>1$ shows, truth is, a qualitatively different
tail-to-body (\emph{i.e. }tail to plateau and fall) response. While
for the case $\text{\ensuremath{\beta}}<1$ the tail have the look
of a natural extension of the fall, in the case $\beta>1$ there is
an abrupt change of direction between the body fall at medium frequencies
and the high frequency tail. We draw on to the representations of
$(\log_{10}|\psi_{\beta}|)'$ and $(\log_{10}|\psi_{\beta}|)''$ to
explain the changes. (See figure \ref{Figure 02}). The first logarithmic
derivative it is no longer a simple step, after the drop a narrow
well is formed, (with the bottom at $\nu\lesssim0.8$), and following
this the line comes up and stabilizes at the selected value of $-\beta$,
the same of the asymptote predicted by the series. What it is remarkable
since the functional series from we obtained the expression for potential
tail it is not converging in this case. (See figure \ref{Figure 02}\emph{a}).

As a direct consequence of this new geometrical characteristic of
the first derivative, the second logarithmic derivative presents two
peaks, one negative and the other positive. The negative is a parametric
continuation of the family shown for the case $\beta<1$, and has
as a new property certain degree of skewness due to the presence of
the second peak. Again the position of the maximum in OX axis grows
in frequency and approaches to $\nu_{max}\lesssim1$. It never reaches
that 'high' value for the interval $1\leq\beta\leq2$, again because
the second peak. The set of second peaks, looked at as a parametric
family, accrues their maxima near to $\nu_{max}\sim1$ as $\beta\rightarrow2$.
They are smaller than their corresponding partners. (See figure \ref{Figure 02},
there in the boxes of the first and second derivatives of $\log_{10}|\psi_{\beta}|$,
for sake of clarity, are not depicted all the curves available but
we have selected the last one of the previous group, $\beta=1$, the
curves corresponding to $\beta=1.x0$ with $x\in$$\{1,2,...,9\}$
and $\text{\ensuremath{\beta}}=2.00$). The first peak corresponds
to the fall of the body of the function $\log_{10}|\psi_{\beta}|$
and this one is even steeper than in the previous curves $\beta<1$
as the depth of the minimum grows in magnitude. The second peak, the
positive, smaller in magnitude than the preceding negative, is a new
phenomenology not seen in the previous case. It indicates a fold upwards
in the curve $\log_{10}|\psi_{\beta}|$ in a very narrow region of
frequencies as the short half-width of the peaks is developed in an
interval with upper limit not beyond $\nu\sim2$ as $\beta\rightarrow2$.
Of course for values of $\beta\lesssim1.40$ the half-width is greater
than this limit value but the height of the peak is so small compared
to greater values of $\beta$ that the sidestep of the curve $\log_{10}|\psi_{\beta}|$
in this region is barely perceptible. (See figure \ref{Figure 02}\emph{b}).
Beyond this range of frequencies (\emph{i.e.} $\nu\geq10$) the 'activity'
of the second derivative, $(\log_{10}|\psi_{\beta}|)''$, is negligible
and the function is mostly of a potential nature. In the figure \ref{Figure 02}\emph{c},
as in the former case a graph of the exponent of tail\emph{ vs. }$\beta$
parameter is shown. We hold the same notation as before, $a_{l}$
for the result of the adjustment of $M\nu^{-a}$ to $\log_{10}|\psi_{\beta}|$,
in the interval $\nu\in(20.0,500.0005]$ and $a_{h}$ for the fit
in the range of $\text{\ensuremath{\nu}}\in(20.0,10^{7}]$. For the
naked eye is quite difficult to distinguish the trend of these two
sets of parameters from the result $a=\beta$, (solid line in the
figure), what it points to an early high frequency tail very different
in behaviour from the main body of the spectrum.

\section{Common approaches}

\subsection{The asymptotic trends of 1-HN approximation}

\subsubsection{The stretched instance $\beta<1$}

We have drawn two conclusions from the asymptotic behaviour of $|\psi_{\beta}|$.
Firstly from a point of view of reconstruction in $t-$space only
an small interval of low frequencies are needed to recover the main
traits of the relaxation. Approximately the most significant part,
--plateau, bend and fall--, happens in the range of $\nu\in[0,1]$
for most the betas and the tail almost stabilizes to its asymptotic
functional form before $\nu<10$. And second this tail performs quite
well as a potential decay of exponent $\beta$, \emph{i.e.} $|\psi_{\beta}|\approx M\nu^{-\beta}$.
Consequently, keeping in mind this two important features, and with
the intention to design the best possible approximation, we shall
test how the Havriliak-Negami function it sticks to the data, $|\psi_{\beta}|$,
when we optimize the error between them. In short we wonder how the
approximation in area distorts the desirable common path of the tails
and reciprocally how the high frequency concomitance of the tails
spoils the very low frequency description of a function, ($|\psi_{\beta}|$),
which is steadier than its Havriliak-Negami alternate.
\begin{figure}
\centering{}\includegraphics[width=0.8\columnwidth]{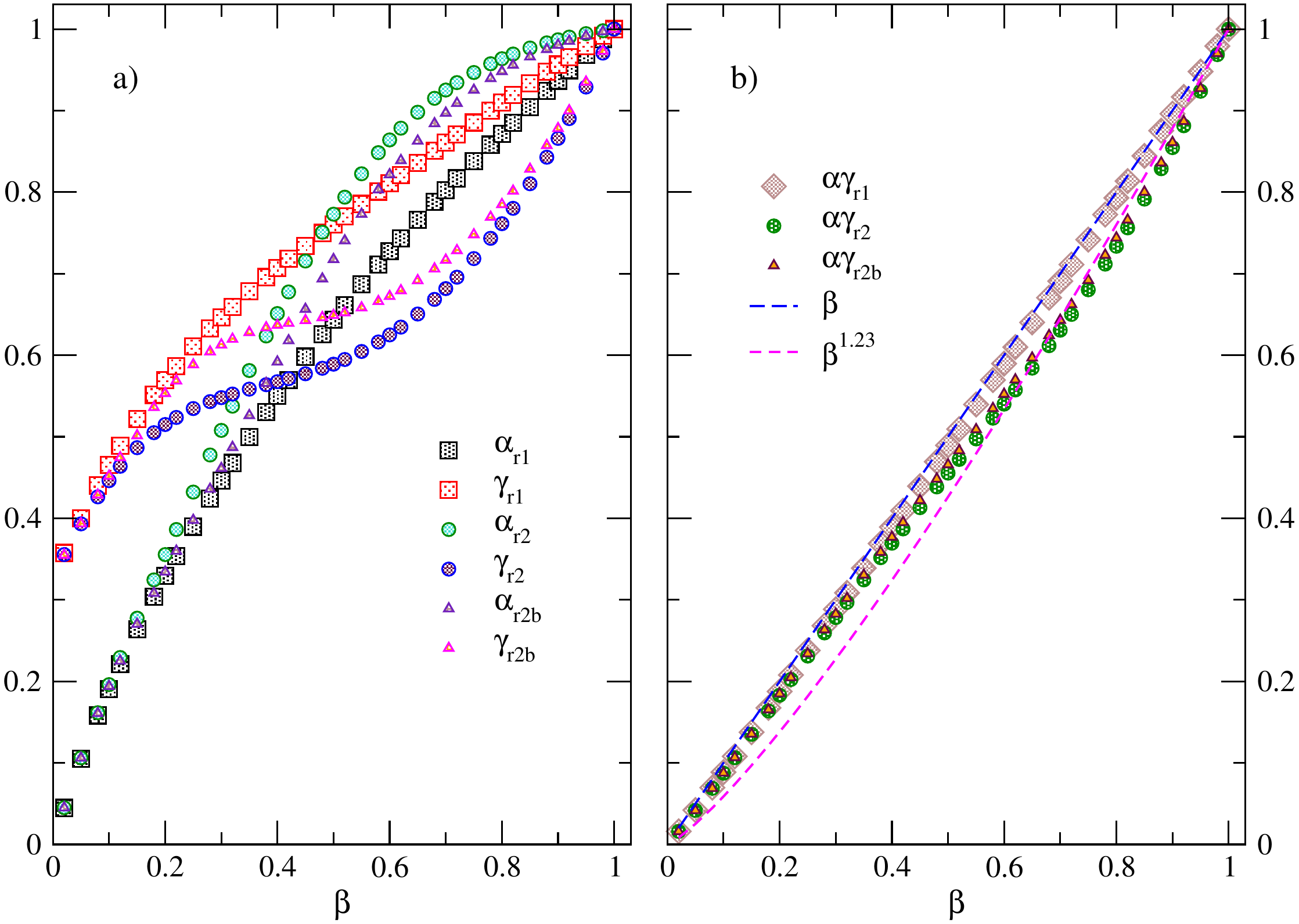}\protect\caption{{\footnotesize{}\label{Figure 03}Left panel, }\emph{\footnotesize{}a}{\footnotesize{}):
Parameters $\alpha$ and $\gamma$ got from adjustment $\psi_{\beta}(\omega)\approx A*HN_{\alpha,\gamma}(\omega)$.
The coarser step sampling, and $A\equiv1$, is for mesh $r_{1}$.
Finer steps are those of $r_{2}$, ($A\equiv1$), and $r_{\text{2b}}$,
$A\protect\neq1$). Right panel, }\emph{\footnotesize{}b}{\footnotesize{}):
Products $\alpha\cdot\gamma$ for the previous parameters. Dashed
lines would correspond to situations $\alpha\cdot\gamma$$=\beta$
and $=\beta^{1.23}$.}}
\end{figure}

To this end we made three related adjustments of the data in the interval
$\nu\in[0,500.0005]$ to one Havriliak-Negami function. In the first
case the sampling step in frequency interval was $\delta\nu=500/999.999$,
that is a purged sample of the points used in the second and third
cases, both with regular stride $\mbox{\ensuremath{\delta\nu}}=1/999.999$.
Thus the influence of the body is practically eliminated and the tail
is the responsible of the values of the parameters obtained for each
beta. In the second case a standard fit is done, and in the third
one the initial condition $HN_{\alpha,\gamma}(\omega=0)=1$ is relaxed
to a free value, \emph{i.e. }$\psi_{\beta}(\omega)\approx$ $A*HN_{\alpha,\gamma}(\omega)$
with $A$ to be determined in the optimization too. The idea behind
this is to allow the Havriliak-Negami approximant compensates its
lack of suitability in the very low frequencies, (there it doesn't
measure up, given its modulus is below $|\psi_{\beta}|$), and to
soften the influence of the tail in relationship to the body part
of data. The result, for $\beta<1$, points to the sensibility of
$\text{\ensuremath{\alpha}}$ and $\gamma$ parameters to small changes
in the approximation of $|\psi_{\beta}|$ area by $A*HN_{\alpha,\gamma}(\omega)$,
--even with so small variations of $A$ as a maximum 6\% around $\beta=0.32$--,
and the sturdiness of tail influence on the asymptotic trend of $HN_{\alpha,\gamma}$.

In figure \ref{Figure 03}, to the left, we depict parameters $\alpha$
and $\gamma$ for the three referred cases, and in the right box the
corresponding products $\alpha\cdot\gamma$ are graphed. In both boxes
the subscripts $r_{1}$ are for the purged case, the $r_{2}$'s are
for the common fit, $\psi_{\beta}\approx HN_{\alpha,\gamma}$, and
the modified case $\psi_{\beta}\approx A*HN_{\alpha,\gamma}$ is denoted
with $r_{2b}$.

We can see how $\alpha_{r1}(\beta)$ and $\gamma_{r1}(\beta)$ have
a monotonous increasing behaviour and the curves only touch themselves
at $\beta=1$. A changing picture when a finer mesh is used and the
very low frequencies are added to the data, now the curves $\alpha_{r2}(\beta)$
and $\gamma_{r2}(\beta)$, still increasing ones that meet at $\beta=1$,
cross one to the other within the interval (approx. at $0.32<\beta<0.35$).
And the same happens to the extended approximation $A*HN_{\alpha,\gamma}$,
though the crossover of $\alpha_{r2b}(\beta)$ and $\gamma_{r2b}(\beta)$
it is shifted to the right of the former ($0.42<\beta<0.45$). The
$\alpha_{r?}(\beta)$ curves don't change qualitatively, as their
all first derivatives keep monotonously decreasing, but the $\gamma_{r?}(\beta)$'s
do inasmuch as the first derivative suffers a transition from strictly
decreasing function, ($r_{1}$), to one with a minimum (\emph{i.e.
}it decreases, halts and then increases for cases $r_{2}$ and $r_{2b}$).
Besides for both families, $\alpha$ and $\gamma$, an apparent transition
from $r_{1}$ to $r_{2}$ is observed in case $r_{2b}$. It suggests
an enhancement of the weight of tail in detriment of plateau and fall
weights, what it sounds logic insofar as the extended condition frees
the constraint of approximating the low frequency zone by the correspondent
region of Havriliak-Negami function, and this allows a greater influence
on the remaining part by the data of high frequency. That is to say
we do not provide to $HN_{\alpha,\gamma}(\omega)$ with a new functional
flexibility, this should be achieved for example with additional terms,
by contrast now the fit is obtained by coating, instead of adapting,
the plateau of data. All this sets free the function tail to adapt
itself to respective data tail. Consequently the appearance of transition
that shows $r_{2b}$, stresses how carefully the design of the approximant,
the method of optimization, the data absent of errors, and of course
the frequency domain, should be done or chosen.

This is also seen in the graph of the tree products $\text{\ensuremath{\alpha}}\cdot\gamma_{r?}(\beta)$,
(figure \ref{Figure 03}, right panel). The first one, $\alpha\cdot\gamma_{r1}$,
follows heavily a straight line $y=\beta$, indeed a regression to
$y=\beta^{a}$ gives $a=1.035$ with a correlation of $c.c.=0.999988$.
The second and third cases ($r_{2}$ and $r_{2b}$), give $a=1.143$
with $c.c.=0.998633$ and $a=1.113$ with $c.c.=0.999004$ respectively.%
\footnote{As a matter of fact a better expression, $1-(1-\beta)^{a}$, is possible
though it would make more difficult the comparison to formulas like
$\beta^{\bar{a}}$ found in the scientific literature.%
} In all three cases the dependency deviates from proposed law $y=\beta^{1.23}$
\cite{Medi 2011,Alva 1991,Alva 1993}, it is not much to say about
$r_{1}$ since it holds data almost exclusively from tails, nevertheless
$r_{2}$ and $r_{2b}$ maintain, by sampling, the information needed
to reconstruct the original function to large times (\emph{i.e. }low
frequencies), and is quite significant that the approaching to the
mentioned law is different according to the value $\beta$. When $0<\beta<0.6$
there is no closeness at all, and if $0.6<\beta<1.0$ the agreement
is only due to the similar functional dependence, $\beta^{a}$. The
point here is, as $\beta$ evolves from 0 to 1 the $(\log_{10}|\psi_{\beta}|)'$
function takes form as an abrupt step which implies for $|\psi_{\beta}|$
a greater width for the plateau and a lesser magnitude and role for
the tail, therefore while adjusting $\psi_{\beta}$ to a Havriliak-Negami
function, a progression from tail dominated, ($y\sim m*\beta$, $m<1$),
to body dominated fit, ($y\sim\beta^{a}$, $a>1$), will take place
in product $\alpha\cdot\gamma(\beta)$. Logically this transition
will be spoiled if the tails for small beta values does not develop
properly and they are rounded off at the limits of an $\omega-$space
finite interval, such thing happens for the FFT and is expected for
an algorithm that erodes sharp characteristics whilst it determines
the density of relaxation times \cite{Medi 2011,Prov 1982,Iman 1988}.
In such circumstances each different procedure used will change the
beta dependence of $\alpha\cdot\gamma$ in a particular and own way
\cite{Schr 2010,Medi 2011,Alva 1991,Alva 1993}.
\begin{figure}
\centering{}\includegraphics[width=0.8\columnwidth]{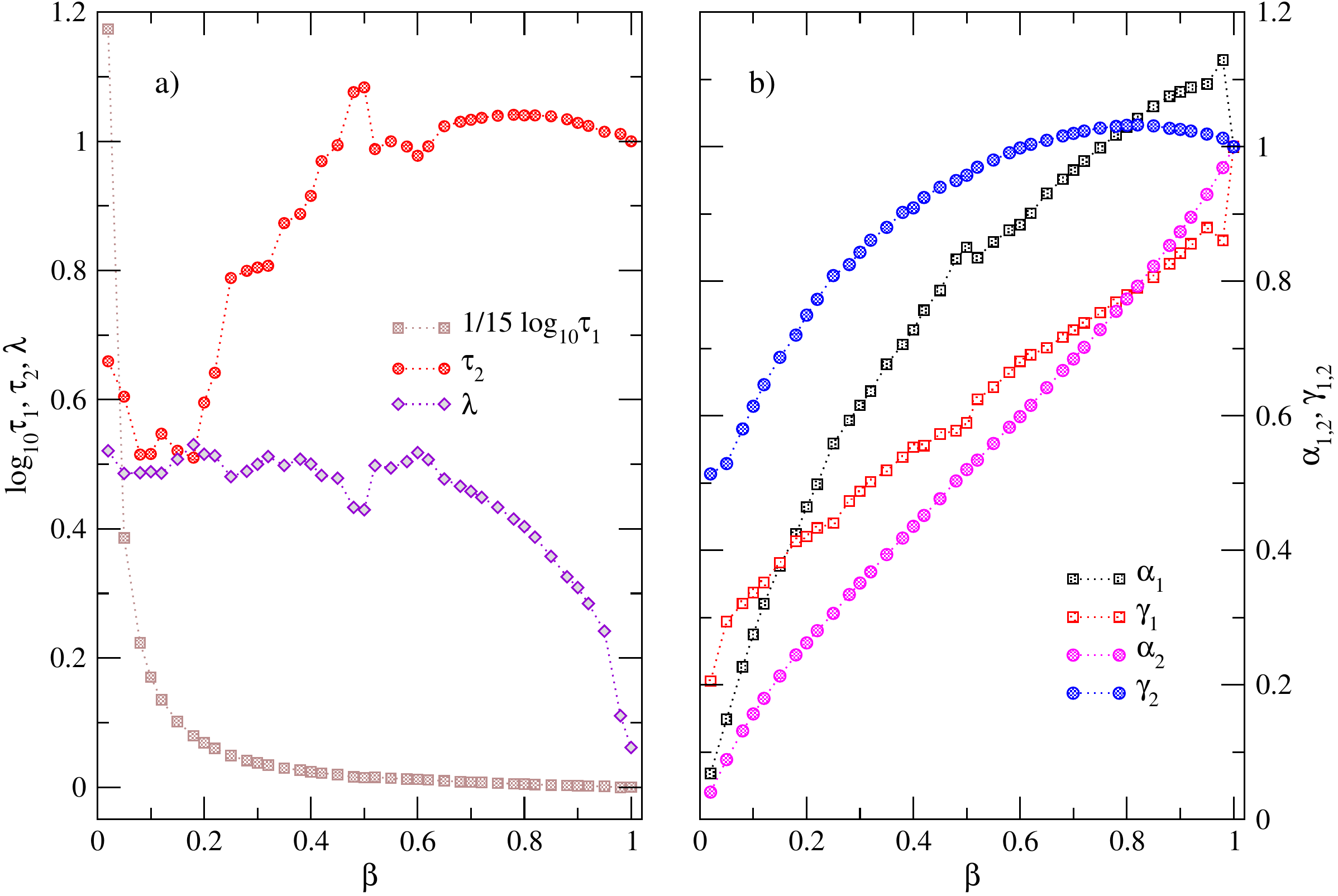}\protect\caption{{\footnotesize{}\label{Figure 04}Parameters, $\tau_{1,2}$ and $\lambda$
in the panel }\emph{\footnotesize{}a}{\footnotesize{}, and in the
}\emph{\footnotesize{}b}{\footnotesize{} one $\alpha_{1,2}$ and $\gamma_{1,2}$,
for an approximant $\mathcal{A}p_{2}HN_{\alpha,\gamma,\tau,\lambda}(\omega)$,
as in Eq. \ref{Eq: 1}, obtained adjusting the function to, ($r_{1}$),
sampled data of $|\psi_{\beta}|$ in the range of medium frequencies.
$\beta\leq1$. Observe the similarities with a fit of same kind for
high frequencies. Only $\lambda$ (medium) shows a behaviour more
akin to low frequencies than to high ones. See figures \ref{Figure 05}
and \ref{Figure 06}. Dotted lines are just an aid for the eye.}}
\end{figure}

\subsubsection{The squeezed instance $\beta>1$}

The linear approach to $\alpha\cdot\gamma(\beta)$ nevertheless fails,
for same kind of samplings $r_{1}$, $r_{2}$ and $r_{2b}$, when
$1<\beta<2$. In this region of control parameter $\beta$ a good
empirical fitting will be $y=\beta$$+A_{1}(\beta-1)^{1.3}$$+A_{2}(\beta-1)^{3.7}$$+A_{3}(\beta-1)^{10}$
with a correlation of $0.999999$ for the three cases. Obviously the
magnitude of parameters will vary being the look of $r_{1}$ different
of $r_{2}$ or $r_{2b}$. While at first sight $r_{1}$ could seem
a deviation from linearity easily corrected with a simple power of
$(\beta-1)$ --it is possible to correct $y\sim\beta$ with a quadratic
power to obtain a good agreement-- the remaining two cases need more
higher power terms to give the impression of a rapidly growing function.
Besides in all cases $A_{i,r1}<<$$A_{i,r2b}<A_{i,r2}$, with $i=1,2,3$
and $A_{2}$ bigger than the other two $A's$, allowing the interpretation
of $r_{2b}$ as an intermediate case between a tail dominant adjustment
and a body dominant one. Exactly as in interval $0<\beta<1$, though
with a great difference, the difficulty to obtain a behaviour that
resembles the asymptotic trend of data $y\approx\beta$.

This hindrance should be borne in mind when making an approximation
to $\psi_{\beta}$. As there is a sudden change of direction in the
modulus from the dropping to the tail around $\nu\approx1$ and a
very fast decreasing of magnitude, any function showing monotony in
decreasing and lack of oscillations or ripples will have large difficulties
to follow the data from low to high frequencies. Therefore our choice,
a description by a simple Havriliak-Negami approximant, won't keep
on track with tail inasmuch as the low frequency part of data is heavier
and prevailing than the high frequency part and the fitting function
too mild for dodging the twist presented by the second logarithmic
derivative. Consequently, even when a residual quantity of data of
such frequencies are present in the sample, the optimization process
will balance towards a selection of parameters describing mainly the
plateau and rise height of step rather than the tail. This is what
happens with case $r_{1}$ which deviates from linearity $y=\beta$
and of course explains, when the whole sample of data is used in $r_{2}$
and $r_{2b}$, the explosion of product $\alpha\cdot\gamma(\beta)$,
for $1<\beta<2$.

\subsection{The asymptotic trends of 2-HN approximation}

While the approximation of $\psi_{\beta}$ by means of one Havriliak-Negami
function describes properly tail decaying of data for $\beta<1$,
--although the asymptotic tendency of approximant suffers distortion
because the matching to plateau zone--, it is clear from scientific
literature that it is not enough to give an account of shape for the
low frequencies \cite{Weis 1994,Medi 2011,Alva 1991,Havr 1996}. Neither
is the case for $\beta>1$ as the advantage of picturing the high
frequency asymptote is lost because tail weight in the optimization
process can not offset plateau and drop predominance. However functional
proximity can be improved adding a new H-N term for adapting to FFT
of finite size temporal data \cite{Macd 1986,Medi 2011}. Is our intention
to show that the same is true for the numerically calculated complex
numbers $\psi_{\beta}(\omega)$ yet with an added feature, a right
portrayal of the tail. Something that was not previously attained
inasmuch as a finite window in the FFT distorted by convolution the
high frequency part of spectrum. Our goal is then to show how this
two terms approximant gives a good picture of both low and high frequencies.
And how each term shares its contribution to the global approximation,
a pertinent question because the roles of both terms will not be equal
for cases $\beta<1$ and $\beta>1$.
\begin{figure}
\centering{}\includegraphics[width=0.8\columnwidth]{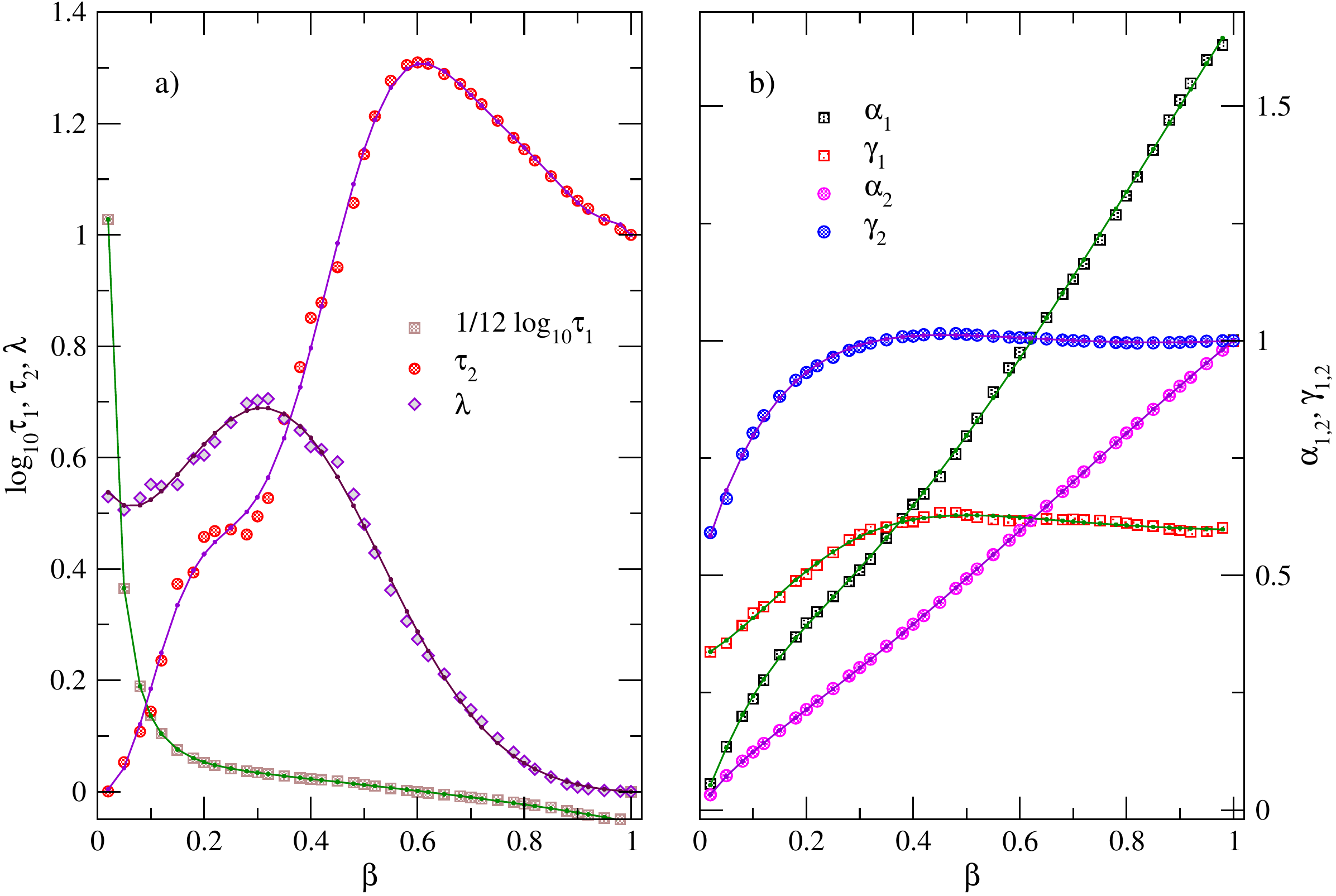}\protect\caption{{\footnotesize{}\label{Figure 05}As in figure \ref{Figure 04} parameters
$\{\alpha_{1,2},\gamma_{1,2},\tau_{1,2},\lambda\}$ are calculated
for a double approximant of Havriliak-Negami type. See Eq. \ref{Eq: 1}.
The frequency of data employed reaches up to $\nu\sim10^{12}$ with
$\beta\leq1$. Lines are mathematical functions designed to describe
these curves as dependent with variable $\beta$. See tables \ref{tab:Tab Ia}
and \ref{tab:Tab Ib} in appendix.}}
\end{figure}

\subsubsection{The stretched case $\beta<1$}

We have used the series of data till $\nu=500.0005$ in both versions
$r_{2}$ (\emph{i.e. $\delta\nu_{r2}=1/999.999$}), and $r_{1}$ ($\delta\nu_{r1}=500*\delta\nu_{r2}$)
and the logarithmically homogeneous one $r_{sl}$ ($\delta\nu_{i,sl}=9*10^{a_{i}-3}$),
until $\mbox{\ensuremath{\nu}}=10^{12}$, or $\mbox{\ensuremath{\nu}}=10^{7}$,
for $\beta<1$ or $\beta>1$, respectively. The two terms approximant
already described which will adjust data is
\begin{equation}
\mathcal{A}p_{2}HN_{\alpha,\gamma,\tau,\lambda}(\omega)=\sum_{s=1}^{2}\frac{\lambda_{s}}{(1+(i\tau_{s}\omega)^{\alpha_{s}})^{\gamma_{s}}}\label{Eq: 1}
\end{equation}
with $\lambda_{1}\equiv\lambda$ and $\lambda_{1}+\lambda_{2}=1$.

\subsubsection*{Mesh sampling (\emph{$r_{1}$})}

In figure \ref{Figure 04} are presented the fitting parameters for
case $r_{1}$ which is dominated by tail values in medium frequencies
as the low frequencies are mainly absent by decimation. The right
panel describes the responsible ones for the potential behaviour of
tails, $\alpha_{1,2}$ and $\gamma_{1,2}$, and the left panel describes
the characteristic times of both functions, $\tau_{1,2}$, plus share
coefficient $\lambda$. In the left panel is convenient to take notice
of a slowing varying lambda in interval $0.02\leq\beta\lesssim0.6$
with a value near to 0.5 which suddenly drops to zero from $\beta\approx$
0.60 to 1.00, also is remarkable the difference in magnitude between
$\tau_{1}$ and $\tau_{2}$. While $\tau_{2}$ never exceeds the value
of 1.1, $\tau_{1}$ requires of a decimal logarithmic scale to be
shown in the same graph. It gives the idea of a dominant function
associated with $\tau_{2}$ only corrected by the one associated with
$\tau_{1}$ in the necessary zone of low frequencies, an oversimplified
picture because is more accurate with values of $\beta$ near to 1.00
than with values near to 0.02 as both functions decay very slowly
in this range of parameter $\beta$, due to small values of $\alpha_{1}$
and $\alpha_{2}$. However it is a good frame of reference to talk
about the contribution of each $HN_{s=1,2}(\omega)$, so we'll refer
recurrently in these terms to them.

In the right panel two features should be noticed, first $\alpha_{2}$
and $\gamma_{2}$ do not cross one each other in line with what we
saw while using only one Havriliak-Negami function to describe pruned
data. And second in consonance with the fact that $\lambda$ diminishes
in value approximating to zero and the numerical optimization is locked
by only one $HN(\omega)$ function, $\alpha_{1}$ and $\gamma_{1}$
both present an apparent discontinuity at $\beta=1$. Although not
all of these two functions is wrong as they cross themselves a property
of $\alpha$ and $\gamma$ already shown in the case of approximating
data, (low frequencies included), with only one Havriliak-Negami function.
This fact enforces the idea of considering $HN_{1}(\omega)$ a backup
for $HN_{2}(\omega)$ while fitting functions to data in low frequency
zone. (See figure \ref{Figure 04}).
\begin{figure}
\centering{}\includegraphics[width=0.8\columnwidth]{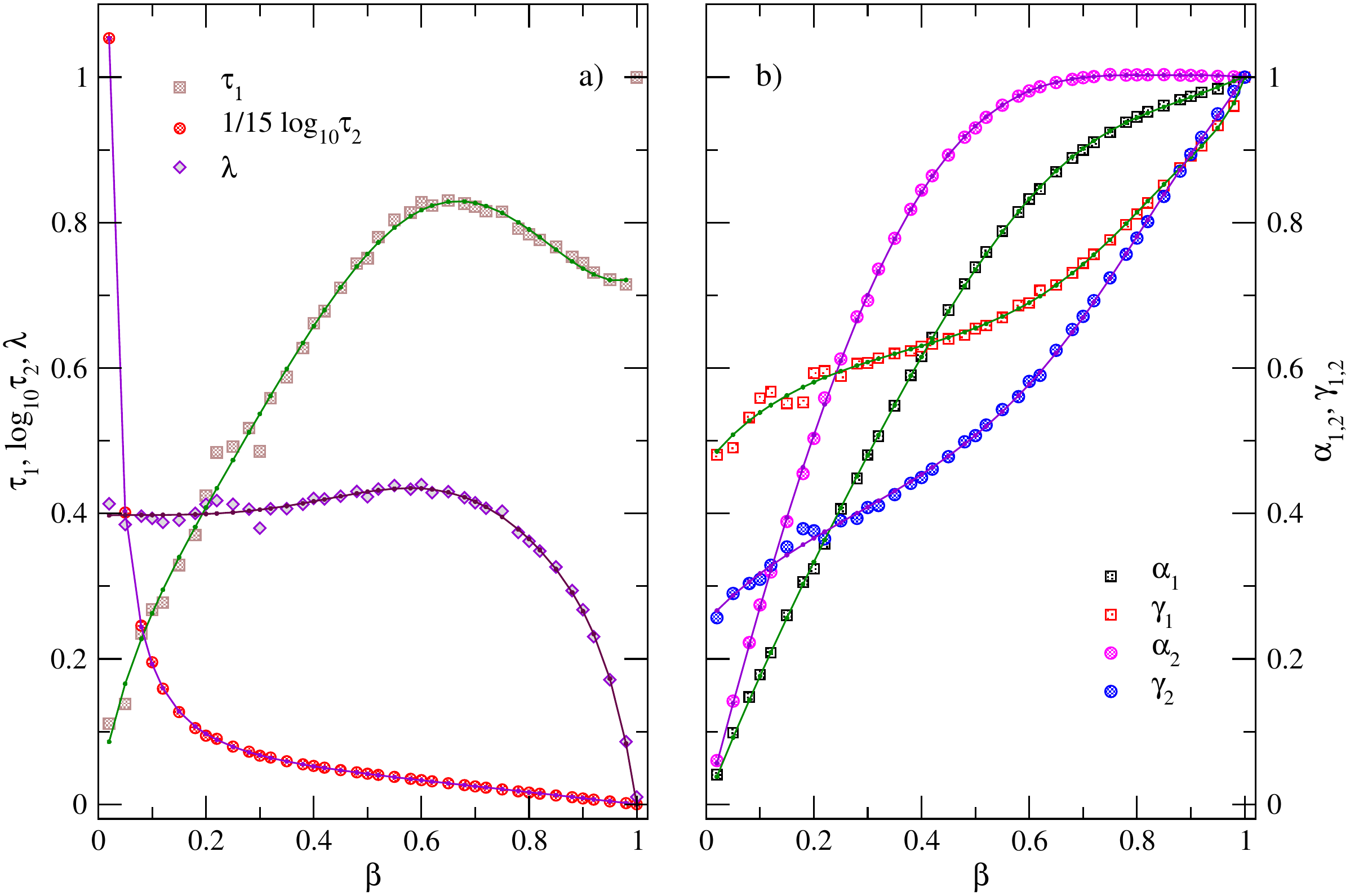}\protect\caption{{\footnotesize{}\label{Figure 06}Parameters $\{\alpha_{1,2},\gamma_{1,2},\tau_{1,2},\lambda\}$
obtained after minimizing $\big||\psi_{\beta}|-|\mathcal{A}p_{2}HN_{\alpha,\gamma,\tau,\lambda}(\omega)|\big|$
in the interval of frequencies $\nu\in[0,500.0005]$, with $\delta\nu=1/999.999$.
$\beta\leq1$. Solid lines are analytical rough estimates to describe
the parameters. See tables \ref{tab:Tab IIa} and \ref{tab:Tab IIb}
in appendix.}}
\end{figure}

\subsubsection*{Mesh sampling ($r_{sl}$)}

With the aim of testing this hypothesis, namely the residuary influence
at low frequency of ancillary Havriliak-Negami function ($HN_{1}(\omega)$),
a further adjustment to data with larger interval of frequencies is
done. Now the sampling called logarithmically homogeneous, $r_{sl}$,
is used as previous data have a very fine step $\delta\nu$, which
makes computationally hard any optimization in a sensible amount of
time. The interval of frequencies is now $\nu\in[1.00,10^{12}]$ for
$\beta<1$ with the aforementioned step $\delta\nu_{i,sl}$ which
changes in each decade of frequency. After calculations a new layout
springs out as result for the parameters of approximant, this is shown
in figure \ref{Figure 05}. We observe important differences, comparing
to latter result, for every parameter. For share coefficient lambda
the quick decreasing starts sooner, nearly $\beta\sim$ 0.4 and has
opposite curvature to the prior case, besides the values attained
in the interval $0<\beta\lesssim0.4$ are comprised between 0.5 and
0.7 and do not seem steady as before. The curve for $\tau_{2}$ also
suffers modifications, now is smoother and starts near zero for $\beta\sim0$
and reaches one for $\beta=1$. This is in consonance with the errors
and the final optimization for $r_{2}$ sampling case and suggests
that the former curve for $r_{1}$ suffers from an strong source of
error, possibly an interference between $HN_{1}(\omega)$ and $HN_{2}(\omega)$
in the initial interval $0.02\leq\beta<0.18$. A hint to explain such
situation is to realize the opposite results of $\tau_{1}$ and $\alpha_{1}$.
While $\tau_{1}$ would imply a very quick decay of $HN_{1}(\omega)$,
the small value of $\alpha_{1}$ would slow down such decay, and would
yield a non negligible contribution, something not to be expected
from greater values of $\alpha_{1}$.

This balance allows a slight rivalry with the values of $HN_{2}(\omega)$
also dominated by a small $\alpha_{2}$. As a consequence of this
equilibrium large fluctuations of $\tau_{2}$ in the neighborhood
of $\beta\sim0$ make sense. Finally we should add that interferences
in the vicinity of $\beta\sim1$ between $HN_{1}(\omega)$ and $HN_{2}(\omega)$
also occur affecting $\tau_{1}$, because for this curve a discontinuity
happens at $\beta=1$ since $\tau_{1}(\beta=1)=1$ and $\tau_{1}(\beta\lesssim1)<1$.
And $\tau_{1}$ is not the only parameter with such discontinuity,
$\alpha_{1}$ and $\gamma_{1}$ also take values quite far from their
ideal amount of 1 near $\beta\lesssim1$. (See figure \ref{Figure 05}).
This pathology, partially consequence of a negligible $\lambda$,
points to a minor and complementary role of $HN_{1}(\omega)$ describing
$\psi_{\beta\lesssim1}(\omega)$ tail. A trend which is the opposite
when the beta values are small, as the change of almost every parameter
curve in the neighborhood of $\beta\sim0.4$ seems to suggest. It
is worthy to note the coincidence of this turnaround with the fact
the tails are not fully developed at very high frequencies for lesser
values of beta than 0.4, at least when a comparison with greater values,
(\emph{v. gr}. $\beta>0.6$), is done. See figure \ref{Figure 01},
graph \emph{a}. To finish we should like to remark the quasi constant
behaviour of $\gamma_{1,2}$ in the interval $0.40\lesssim\beta<1$
and the quasilinear one of $\alpha_{1,2}$ in the whole interval of
beta, making the relaxation of tails for $HN_{2}(\omega)$ nearer
to a Cole-Cole type, ($\gamma\equiv1$), than to a Cole-Davidson one,
($\alpha\equiv1$). By contrast relaxation for $HN_{1}(\omega)$ still
remains Havriliak-Negami in both cases $r_{1}$ and $r_{sl}$ though
they are very different as a family of curves, (compare $\gamma_{1}$
in figures \ref{Figure 04} and \ref{Figure 05}), inasmuch as $\gamma_{1}$
in $r_{1}$ is an increasing function for all the beta domain and
$\gamma_{1}$ in $r_{sl}$ holds itself as quasi constant in the mentioned
subinterval $\beta\in(0.4,1.0]$. Surely the lesser weight of medium
and low frequencies in $r_{sl}$ with respect to $r_{1}$ takes its
toll.

\subsubsection*{Mesh sampling ($r_{2}$)}

Finally an adjustment to an $r_{2}$ sampling which includes many
points corresponding to low frequencies is done, the result is that
the roles of characteristics times $\tau_{1}$ and $\tau_{2}$ are
interchanged. Now $\tau_{1}$ scale does not surpass the value of
one. For betas near to one $\tau_{1}$ is around 0.7 and at $\beta=1$
is discontinuous, besides after a maximum located at $\beta\approx0.62$
it decreases to 0.1 as $\beta\rightarrow0.02$. Meanwhile $\tau_{2}$
is a monotonous decreasing curve from $\beta\sim0$ to $\beta=1.00$
which can not be represented in the same scale as before unless decimal
logarithms are taken and rescaling is done. (See left panel in figure
\ref{Figure 06}). This suggests for the whole range of beta two scenarios,
one in which the first term of the approximant, $HN_{1}(\omega)$,
losses progressively importance as beta grows. It starts at $\beta\approx0.62$
and ends when the approximant becomes degenerate $\mathcal{A}p_{2}HN_{\alpha,\gamma,\tau,\lambda\rightarrow0}(\omega)\approx HN_{\{\alpha_{2},\gamma_{2},\tau_{2}\}\rightarrow\{1,1,1\}}(\omega)$,
at $\beta=1.00$, (\emph{i.e. }tends to a Debye relaxation as $\lambda\rightarrow0$).
This feature avoids the proper numerical determination of all parameters
of $HN_{1}(\omega)$ and explain why $\tau_{1}$ is discontinuous,
the optimization is locked by the suitability of $HN_{2}(\omega)$
and no further contribution of $HN_{1}(\omega)$ is possible. On the
contrary, in the second scenario, ($0.02\leq\text{\ensuremath{\beta}}<0.62$),
lambda, which takes values near to 0.4, shares out equal amount of
contribution to each Havriliak-Negami term. Now the difference in
contribution between both terms is due to the very distinct scale
of time for $\tau_{1}$ and $\tau_{2}$, making one to be the complement
of the other in this fit, while at the same time the tail asymptotic
behaviour is mostly preserved. This situation allows to the approximant
$\mathcal{A}p_{2}HN_{\alpha,\gamma,\tau,\lambda}(\omega)$ a description
for a wide range of frequencies without excessive gap with data, $\psi_{\beta}(\omega)$.
(See figure \ref{Figure 09}). Pairs $\alpha_{1},\gamma_{1}$ and
$\alpha_{2},\gamma_{2}$ show a similar behaviour, --for each pair,
the curves cross themselves and have similar shapes to the case of
one Havriliak-Negami approximant (see figure \ref{Figure 03})--,
what is also consistent with a detailed description of plateau and
drop, (\emph{i.e. }the low frequencies), by jointly both Havriliak-Negami
functions. (See right panel in figure \ref{Figure 06}).
\begin{figure}
\centering{}\includegraphics[width=0.8\columnwidth]{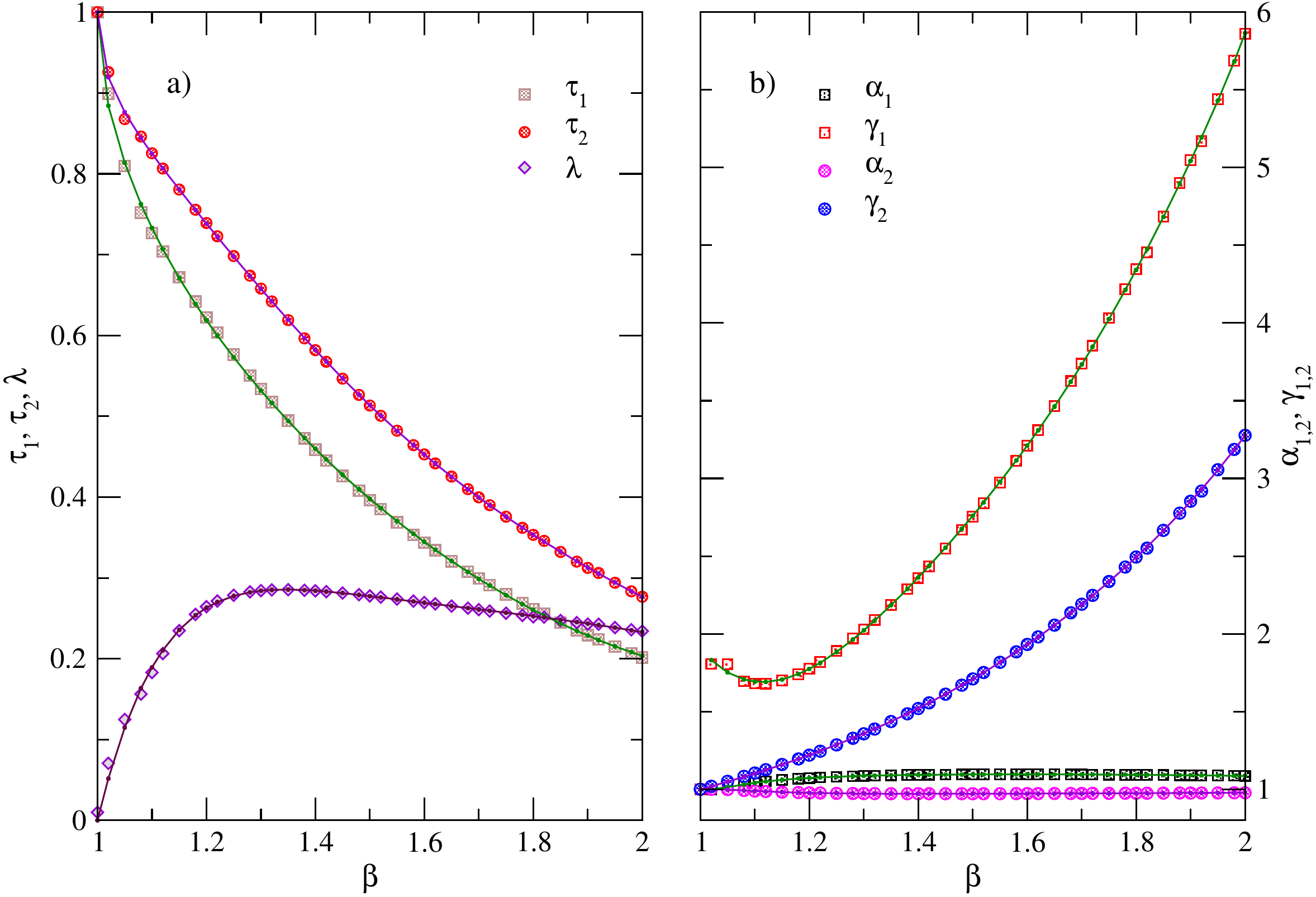}\protect\caption{{\footnotesize{}\label{Figure 07}Parametric curves $\{\alpha_{1,2},\gamma_{1,2},\tau_{1,2},\lambda\}$
of a two-terms approximant, (see Eq. \ref{Eq: 1}), for data, $|\psi_{\beta}|$,
picked from interval $\nu\in[0,500.0005]$ with $\delta\nu=1/999.999$.
$\beta\geq1$. Solid lines are mathematical expressions tailored to
adjust the curves. See tables \ref{tab:Tab IIIa} and \ref{tab:Tab IIIb}
in appendix.}}
\end{figure}

The empirical functions for the adjusting parameters $\alpha_{1,2}$,
$\gamma_{1,2}$, $\tau_{1,2}$ and $\lambda$ are given in tables
\ref{tab:Tab Ia} and \ref{tab:Tab Ib} for long frequency tails ($r_{sl}$),
and in tables \ref{tab:Tab IIa} and \ref{tab:Tab IIb} for low to
medium frequencies ($r_{2}$), all of them are written for interval
$0<\beta\leq1$ of control parameter.

\subsubsection{The squeezed case $\beta>1$}

However a different kind of results are obtained for the interval
$1<\beta\leq2$. We already pointed out how a sudden change in direction
makes bend upwards the curve $|\psi_{\beta}(\omega)|$, which stabilizes
in a potential tail at medium frequencies, after a very sharp dropping
in the low to medium range of them. As a result the transition from
plateau and rise height to tail is marked clearly by the decimal logarithmic
curvature --which presents a chasm and a peak--, and consequently
the adjustment to data by a Havriliak-Negami approximant is made more
difficult, as the latter does not show such an extreme changes in
curvature. The weight of low frequencies is bigger than the data corresponding
to tail because their respective magnitude and total number of points,
(for sampling $r_{2}$), thus the parameters obtained after optimization
are those of a quick decaying function not following the natural asymptote.%
\footnote{Due to the commented features of the function $|\psi_{\beta}|$ near
$\nu\sim0.9$ the influence of clean data in the adjustment to a Havriliak-Negami
approximant it is not very different of same fit to spoiled, (by FFT),
data. In consequence the shape of empirical parameter curves do not
change significantly in either, theoretical or FFT, case. See Figure
\ref{Figure 07}.%
} (See figure \ref{Figure 07}). Accordingly a global description of
data would require necessarily a different two terms approximant of
Havriliak-Negami type for the tail and every $\beta$. To obtain this
we would proceed as before with an additional fit to a sampling like
$r_{1}$ or $r_{sl}$ as both are pruned of low frequencies. In these
cases they are sets of data weighting mostly the medium to high frequencies
and the corresponding result will describe the tail asymptotic behaviour
properly at the expense of not describing at all the plateau and its
environs. Nevertheless the analysis $\alpha_{1,2}$, $\gamma_{1,2}$,
$\tau_{1,2}$ and $\lambda$ as empirical functions of $\beta$ deals
with remnant information of this neglected important part of the spectrum.
\begin{figure}
\centering{}\includegraphics[width=0.8\columnwidth]{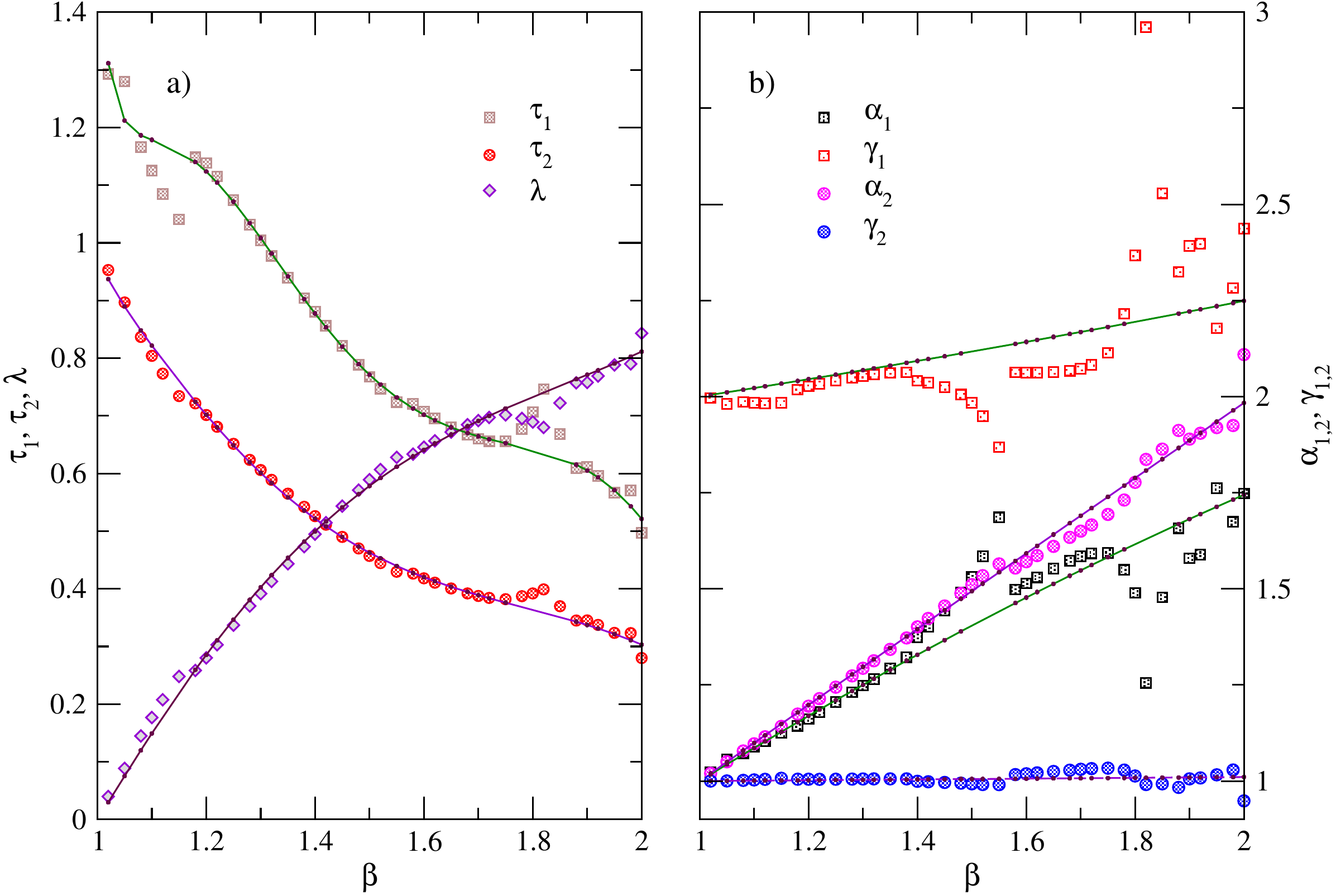}\protect\caption{{\footnotesize{}\label{Figure 08}Parametric curves $\{\alpha_{1,2},\gamma_{1,2},\tau_{1,2},\lambda\}$
of an $\mathcal{A}p_{2}HN_{\alpha,\gamma,\tau,\lambda}$ function
approximating data $|\psi_{\beta}|$ up to $\nu\sim10^{7}$, in a
homogeneous-like logarithmic pace. (See Eq. \ref{Eq: 1} and text).
Notice the big dispersion of experimental points $\alpha_{1}$ and
$\gamma_{1}$, and the correlated corresponding error of $\alpha_{2}$
and $\gamma_{2}$. It is remarkable that, even with this scattering
of the parameters, the products $\alpha_{1}\cdot\gamma_{1}$ and $\alpha_{2}\cdot\gamma_{2}$
hold themselves extremely attached to a linear trend. (See figure
\ref{Figure 09}}\emph{\footnotesize{}c}{\footnotesize{} and }\emph{\footnotesize{}d}{\footnotesize{}).
Solid lines are rough proposals based on constraints given by same
products. See tables \ref{tab:Tab IVa} and \ref{tab:Tab IVb} in
appendix.}}
\end{figure}
 The triplet \{$\alpha_{2}$, $\gamma_{2}$, $\tau_{2}$\} is in both
samplings, $r_{1}$ and $r_{sl}$, describable by means of smooth
functions (\emph{v.gr. }combination of exponentials and polynomials)
and also is the case, except for some occasional dimple in $r_{sl}$,
for the share coefficient $\lambda$. (See figure \ref{Figure 08}).
Nonetheless it is not such a way for the triplet \{$\alpha_{1}$,
$\gamma_{1}$, $\tau_{1}$\}. Though it could be possible to adjust
$\tau_{1}$ with just one mathematical expression and the errors with
data points wouldn't be large, it won't be the case for $\alpha_{1}$
and $\gamma_{1}$. (See right panel in figure \ref{Figure 08}). These
last two curves are crooked, or even broken, in many points and do
not show a defined trend, besides this happens for both cases $r_{1}$
and $r_{sl}$ but each in their own ways. Thus only a piecewise description
is admissible for them. However this is not the treatment we are looking
for the parameters as it would require an interpolating table, and
certainly many errors, instead a simple formula. But not all is lost,
paradoxically the product $\alpha_{1}\cdot\gamma_{1}$ is well behaved
and can be fairly approximated to $\alpha_{1}\cdot\gamma_{1}\simeq2\beta$
in case $r_{sl}$ and to $\alpha_{1}\cdot\gamma_{1}\simeq$ $-0.22+2.06\beta$
in case $r_{1}$. And meanwhile the product $\alpha_{2}\cdot\gamma_{2}$
is even closer and more correlated to a straight line, (\emph{i.e.
}$\alpha_{2}\cdot\gamma_{2}\simeq\beta$ with very small errors in
both cases), than the previous product. The evident conclusion is
that in spite of the bad conditioning and many errors finding $\alpha_{1}$
and $\gamma_{1}$ out the asymptotic tendency of tails is well described
with a two-term Havriliak-Negami sum. On one side the second term,
$HN_{2}(\omega)$, dictates the tail behaviour which at large $\nu$
coincides with the trend of real data, and on the other hand the first
term, $HN_{1}(\omega)$, --quickly fading--, gives a proper placement
of the curve near such data. And for this last constituent, notwithstanding
the poorly defined trends of $\alpha-$ and $\gamma-$type parameters,
the stability of constraint $\alpha_{1}\cdot\gamma_{1}\simeq2\beta$
allows us to put forward an 'ansatz' for their dependence on $\beta$,
which finally enables a finer classification of high frequency spectral
data. This will namely: $\alpha_{1,2}\sim\mathcal{O}(\beta)$, $\gamma_{1}\sim\mathcal{O}(2)$
and $\gamma_{2}\sim\mathcal{O}(1)$.

\begin{figure}
\centering{}\includegraphics[width=0.8\columnwidth]{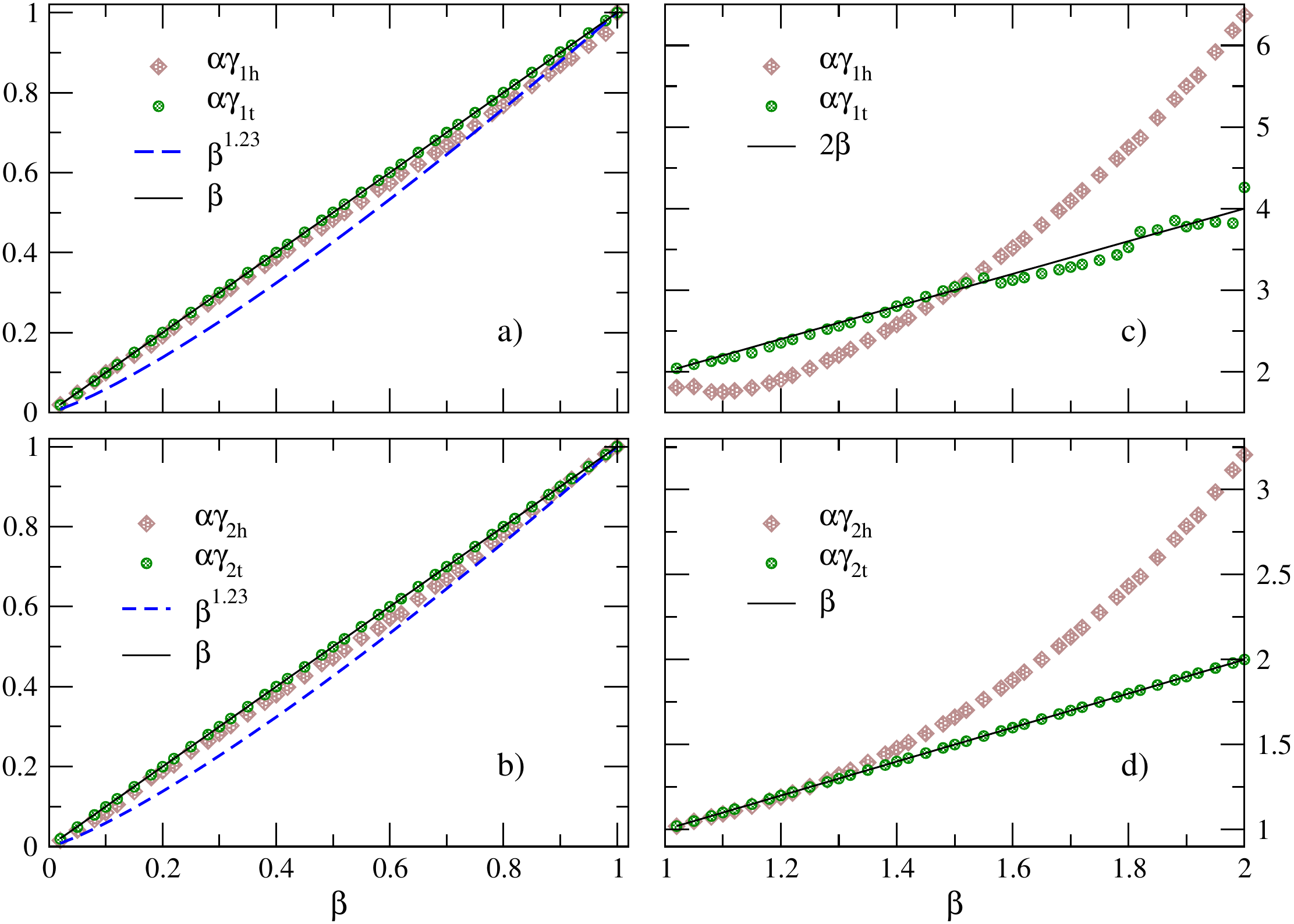}\protect\caption{{\footnotesize{}\label{Figure 09}Products $\alpha_{1}\cdot\gamma_{1}$
and $\alpha_{2}\cdot\gamma_{2}$ of the asymptotic parameters, gotten
by optimizing $\mathcal{A}p_{2}HN_{\alpha,\gamma,\tau,\lambda}\sim\psi_{\beta}$
in samplings $r_{2}$, --subindex $h$: }\emph{\footnotesize{}head}{\footnotesize{}--,
and $r_{sl}$, --subindex $t$: }\emph{\footnotesize{}tail}{\footnotesize{}--,
along the whole interval of variable $0<\beta\leq2$. The solid line
inside boxes }\emph{\footnotesize{}a}{\footnotesize{}, }\emph{\footnotesize{}b
}{\footnotesize{}and }\emph{\footnotesize{}d }{\footnotesize{}represents
$\alpha\cdot\gamma=\beta$, while it is for $\alpha\cdot\gamma=2\beta$
in box }\emph{\footnotesize{}c}{\footnotesize{}. In boxes }\emph{\footnotesize{}a}{\footnotesize{}
and }\emph{\footnotesize{}b}{\footnotesize{} is also depicted the
tendency $\alpha\cdot\gamma=\beta^{1.23}$ with a dashed line.}}
\end{figure}

In Figure \ref{Figure 09} all $\alpha\cdot\gamma$ products already
commented for $0<\beta\leq2$ are depicted. Subscripts numbers $1$
and $2$ stand for the first and second $HN_{?}(\omega)$ term in
each approximant, and letters $h$ and $t$ describe the adjustment
to $r_{2}$ and $r_{sl}$ samplings, respectively.%
\footnote{Denoting $h$ the head, \emph{i.e. }plateau and drop, and $t$ the
high frequency tail, \emph{i.e.} potential asymptotic trend.%
}

Also in tables \ref{tab:Tab IIIa}, \ref{tab:Tab IIIb}, \ref{tab:Tab IVa}
and \ref{tab:Tab IVb} mathematical adjustments to parameters $\alpha_{1,2}$,
$\gamma_{1,2}$, $\tau_{1,2}$ and $\lambda$ are recorded for $1<\beta\leq2$.
In table \ref{tab:Tab IIIa} and table \ref{tab:Tab IIIb} the formulae
represent a quite close match to the empirical points which define
the approximant for low frequencies, the counterpart for high frequencies
is in tables \ref{tab:Tab IVa} and \ref{tab:Tab IVb}. However in
these last two tables the intended formulas for point curves $\alpha_{1}$,
$\gamma_{1}$ and $\tau_{1}$ are just rough estimates as such curves
show breaks and outliers. The remaining point graphs, (\emph{i.e.
}$\alpha_{2}$, $\gamma_{2}$, $\tau_{2}$ and $\lambda$), exhibit
nevertheless an appropriate fit and as a consequence the mathematical
expressions which substitute them give a good description of the high
frequency behaviour (\emph{i.e.} the asymptotes) although possibly
with a misplacement due to errors describing the set of parameters
\{$\alpha_{1}$, $\gamma_{1}$, $\tau_{1}$\}.

\section{Discussion}

By comparing the parametric curves $\{\alpha_{1,2},\gamma_{1,2},\tau_{1,2},\lambda\}$,
(see figure \ref{Figure 06}), with those of a previous approximation
to data $\psi_{\beta}(\omega)$ which undergo convolution with a \texttt{sinc}
function, one can see a strong distortion in curves $\alpha_{1,2}$
and $\gamma_{1,2}$ for low values of $\beta\leq1$. (See figures
3 and 4 in reference \cite{Medi 2011}). In that case they didn't
tend to zero as it happens in our present situation for $\alpha_{1,2}$,
nor retained a value below 1 for $\gamma_{1}$, whithin an interval
of small values of beta ($\beta<0.25$). And of course neither of
them allowed us to consider constraints of type $(\alpha\cdot\gamma)_{1,2}\sim\mathcal{O}(\beta)$
as valid ones. However a successful reconstruction of the Kohlrausch
function was still possible by means of inverse FFT for small ($\beta=0.15$)
and large ($\beta=0.85$) values, despite these facts. 

Some circumstances played a role to compensate the until then unknown
distortion of beta-parametrized curves $\{\alpha,\gamma\}_{1,2}$.
Among them: the serious problems in the medium-to-high frequencies
interval, $\nu\in[10,500]$, were overcome easily with a proper damping
of phase in the approximation, while the problems in low frequency
range --consequence of discrepancy between data and the approximant
pair-- were shadowed by the finite size of time window through the
smearing of peaks. Also the tails of both functions differed so slightly
in a finite interval that made imperceptible the disagreements in
reconstruction at short time scale.

Nevertheless what it was really good for the purpose of ignoring the
distance in function space between the Fast Fourier Transform and
the double Havriliak-Negami approximant, --i.e. its adaptability thanks
to the many parameters--, played against a proper description of the
theoretical Fourier Transform. The main three barriers cited which
oppose to the perfect fit of HN approximant and FT data, will suffer
dissimilar transformation through the process of convolution of data
with a finite window, by the simple reason of being differently developed
head, drop and tail along the same frequency interval. So, the change
of shape induced by convolution, not only in plateaux but also in
the tails where the potential behavior $\omega^{-\beta}$ was rounded
due to finite size of spectrum, will be different with the value of
$\beta$.

In consequence the functional dependence of $\{(\alpha,\gamma,\tau)_{1,2},\lambda\}(\beta)$
for the HN functions has to result distorted with respect to the desired
result of same family while follows theoretical FT data. Such deviation
will depend on $\beta$ inasmuch as the distortion of the FFT is not
equally distributed with this parametric variable. Obviously a lesser
drop tail-to-head in the functions with small $\beta$ values wore
a greater adaptation of the approximants to the convolved data, loosing
some of the original characteristics of theoretical data.

\section{Conclusions}

It is common practice while studying complex systems to analyze their
relaxations in time as well as in frequency. Unfortunately there are
not often at hand short and compact expressions corresponding simultaneously
to the mathematical formulation of a same phenomenon in both spaces.
This article is focused towards the approximation of Fourier Transform
of certain Weibull distributions by Havriliak-Negami functions, studying
their fitting instabilities and suitability. As the suggested models
possess many parameters this first part delves thoroughly into the
functional dependence of them with shape parameter $\beta$. Also
we discuss the influence of data variation and extension on the resulting
$\beta$ dependency, \emph{i. e.} the sensibility of fit with data
errors or distortions. Finally we will end this part establishing
additional restrictions on parameter behaviour which are deduced from
asymptotic trends of the theoretical functions. 

When the values of $\psi_{\beta}(\omega)$ are sampled properly as
to balance the weight of low frequency data with the more numerous
and tiny values of high frequency tails during a fit or optimization,
and besides such data are not distorted or biased by finite time windows
in the Fourier transform of the original function $-\frac{d\exp(-t^{\beta})}{dt}$
or other numerical artifacts as consequence of switching from time
to frequency space \cite{Medi 2011,Alva 1991,Prov 1982,Iman 1988},
it will be true that a unique Havriliak-Negami function shall not
follow other asymptotic condition $\alpha\cdot\gamma=\beta^{d}$ different
of $d=1$.

Also whether two Havriliak-Negami functions are used as approximant
the asymptotic constraints will be $(\alpha\cdot\gamma)_{1,2}=\beta$,
for $\beta\leq1$. While $(\alpha\cdot\gamma)_{1}\approx2\beta$ and
$(\alpha\cdot\gamma)_{2}=\beta$, for $2\geq\beta>1$. Nevertheless
whenever the heads, ($\nu<1$), are involved and the tails, ($\nu>1$),
are not sufficiently important to talk about a fully developed asymptotic
trend, (see figures \ref{Figure 01} and \ref{Figure 02}), the constraints
for $\alpha\cdot\gamma$ won't be fulfilled. In such cases $(\alpha\cdot\gamma)_{1,2}\sim\mathcal{O}(\beta)$
for $\beta\leq1$, and $(\alpha\cdot\gamma)_{1,2}\sim\mathcal{O}(\beta^{d})$
with $d\geq2$ for $\beta>1$.

These constraints should clear up our understanding of the fitting
impediments found at low frequency while the model is being built.
They suggest for slight changes in the uniform model of an approximant
with two terms to give account of the whole range of frequencies.
Unfortunately the accuracy of such approximations is restricted to
the description of modulus of the objective function as far as our
primary purpose is to understand the modifications required over HN
functions to fit data in a better way. An additional work to fully
optimize our models in the complex domain should be done by means
of the information obtained from the new family of relaxation functions
we have introduced here. Work is in progress towards such direction.

\section*{Appendix}

Here are given the tables with formulae and parameter values corresponding
to mathematical adjustments of point data in figures \ref{Figure 05}
to \ref{Figure 08}.

\subsection*{Case $0<\beta\le1$}

\begin{table}
\noindent \begin{centering}
{\tiny{}}%
\begin{tabular}{ccc}
\hline 
\emph{\tiny{}Formulae} & {\tiny{}$\beta A\exp\big[(\sum_{s=1}^{3}a_{s}(1-\beta)^{s})\exp(-M\beta^{d})\big]$} & {\tiny{}$\frac{1}{A\exp\big[(\sum_{s=1}^{3}a_{s}(1-\beta)^{s})\exp(-M\beta^{d})\big]}$}\tabularnewline
\hline 
\hline 
{\scriptsize{}}%
\begin{tabular}{c}
{\scriptsize{}$\frac{Parameters}{Constants}$}\tabularnewline
{\scriptsize{}$A$}\tabularnewline
\multicolumn{1}{c}{{\scriptsize{}$M$ }}\tabularnewline
{\scriptsize{}$d$}\tabularnewline
{\scriptsize{}$a_{1}$}\tabularnewline
{\scriptsize{}$a_{2}$}\tabularnewline
{\scriptsize{}$a_{3}$}\tabularnewline
{\scriptsize{}Corr.}\tabularnewline
\end{tabular} & {\footnotesize{}}%
\begin{tabular}{c|c}
\multicolumn{1}{c}{{\footnotesize{}$\alpha_{1}$}} & {\footnotesize{}$\alpha_{2}$}\tabularnewline
{\footnotesize{}1.68125} & {\footnotesize{}$\equiv$1.000}\tabularnewline
{\footnotesize{}6.00482} & {\footnotesize{}2.14961}\tabularnewline
{\footnotesize{}0.712224} & {\footnotesize{}0.580585}\tabularnewline
{\footnotesize{}-30.8645} & {\footnotesize{}0.56705}\tabularnewline
{\footnotesize{}75.8761} & {\footnotesize{}-2.83736}\tabularnewline
{\footnotesize{}-44.5841} & {\footnotesize{}2.97408}\tabularnewline
\hline 
\multicolumn{1}{c}{\emph{\footnotesize{}0.999858}} & \emph{\footnotesize{}0.999997}\tabularnewline
\end{tabular} & {\footnotesize{}}%
\begin{tabular}{c|c}
\multicolumn{1}{c}{{\footnotesize{}$\gamma_{1}$}} & {\footnotesize{}$\gamma_{2}$}\tabularnewline
{\footnotesize{}1.6738} & {\footnotesize{}$\equiv$1.000}\tabularnewline
{\footnotesize{}4.95993} & {\footnotesize{}2.49211}\tabularnewline
{\footnotesize{}1.17881} & {\footnotesize{}0.651276}\tabularnewline
{\footnotesize{}-5.87091} & {\footnotesize{}0.585986}\tabularnewline
{\footnotesize{}13.4251} & {\footnotesize{}-2.92872}\tabularnewline
{\footnotesize{}-6.94891} & {\footnotesize{}3.0715}\tabularnewline
\hline 
\multicolumn{1}{c}{\emph{\footnotesize{}0.997978}} & \emph{\footnotesize{}0.999256}\tabularnewline
\end{tabular}\tabularnewline
\hline 
\end{tabular}
\par\end{centering}{\tiny \par}

\protect\caption{{\footnotesize{}\label{tab:Tab Ia}}\emph{\footnotesize{}Approximant
of tails, case }{\footnotesize{}$\beta\leq1$: Formulas used to adjust
$\{\alpha_{1,2},\gamma_{1,2}\}$ and their fitting constants \cite{Turn 1998}.
Corr. = correlation coefficient of nonlinear fit. For $\alpha_{1}$
and $\gamma_{1}$, $\text{\ensuremath{\beta}=1}$ is an outlier. See
right panel at figure \ref{Figure 05}.}}
\end{table}
Tables \ref{tab:Tab Ia} and 
\begin{table}
\noindent \begin{centering}
\begin{tabular}{ccc}
\hline 
{\tiny{}}%
\begin{tabular}{c}
{\tiny{}$-0.65664+\dots$}\tabularnewline
{\tiny{}$+\sum_{s=1}^{2}\big\{ A_{s}(1-\beta)+B_{s}(1-\beta)^{2}\big\}\exp(-M_{s}\beta^{0.2})$}\tabularnewline
\end{tabular} & {\tiny{}$\beta^{3}+\beta^{d}\sum_{s=1}^{7}b_{s}(1-\beta)^{s}$} & {\tiny{}$\exp(-M\beta^{d})\sum_{s=1}^{5}c_{s}(1-\beta)^{s}$}\tabularnewline
\hline 
\hline 
{\footnotesize{}}%
\begin{tabular}{c|c}
\multicolumn{1}{c}{{\footnotesize{}$\frac{Parameter}{Constants}$}} & {\footnotesize{}$\log_{10}\tau_{1}$}\tabularnewline
{\footnotesize{}$A_{1}$} & {\footnotesize{}-22.4446}\tabularnewline
{\footnotesize{}$B_{1}$} & {\footnotesize{}-6.64276}\tabularnewline
{\footnotesize{}$M_{1}$} & {\footnotesize{}-0.233028}\tabularnewline
{\footnotesize{}$A_{2}$} & {\footnotesize{}8231.54}\tabularnewline
{\footnotesize{}$B_{2}$} & {\footnotesize{}-7799.77}\tabularnewline
{\footnotesize{}$M_{2}$} & {\footnotesize{}5.59666}\tabularnewline
\hline 
\multicolumn{1}{c}{{\footnotesize{}Corr.}} & \emph{\footnotesize{}0.999984}\tabularnewline
\end{tabular} & {\footnotesize{}}%
\begin{tabular}{c|c}
\multicolumn{1}{c}{{\footnotesize{}$\frac{Parameter}{Constants}$}} & {\footnotesize{}$\tau_{2}$}\tabularnewline
{\footnotesize{}$d$} & {\footnotesize{}2.96596}\tabularnewline
{\footnotesize{}$b_{1}$} & {\footnotesize{}4.54749}\tabularnewline
{\footnotesize{}$b_{2}$} & {\footnotesize{}-32.4329}\tabularnewline
{\footnotesize{}$b_{3}$} & {\footnotesize{}552.577}\tabularnewline
{\footnotesize{}$b_{4}$} & {\footnotesize{}-3159.47}\tabularnewline
{\footnotesize{}$b_{5}$} & {\footnotesize{}9513.52}\tabularnewline
{\footnotesize{}$b_{6}$} & {\footnotesize{}-13492.1}\tabularnewline
{\footnotesize{}$b_{7}$} & {\footnotesize{}7139.92}\tabularnewline
\hline 
\multicolumn{1}{c}{{\footnotesize{}Corr.}} & \emph{\footnotesize{}0.998613}\tabularnewline
\end{tabular} & {\footnotesize{}}%
\begin{tabular}{c|c}
\multicolumn{1}{c}{{\footnotesize{}$\frac{Parameter}{Constants}$}} & {\footnotesize{}$\lambda$}\tabularnewline
{\footnotesize{}$M$} & {\footnotesize{}4.14933}\tabularnewline
{\footnotesize{}$d$} & {\footnotesize{}3.1579}\tabularnewline
{\footnotesize{}$c_{1}$} & {\footnotesize{}3.43714}\tabularnewline
{\footnotesize{}$c_{2}$} & {\footnotesize{}-12.6851}\tabularnewline
{\footnotesize{}$c_{3}$} & {\footnotesize{}35.8224}\tabularnewline
{\footnotesize{}$c_{4}$} & {\footnotesize{}-46.5211}\tabularnewline
{\footnotesize{}$c_{5}$} & {\footnotesize{}20.5172}\tabularnewline
\hline 
\multicolumn{1}{c}{{\footnotesize{}Corr.}} & \emph{\footnotesize{}0.998905}\tabularnewline
\end{tabular}\tabularnewline
\hline 
\end{tabular}
\par\end{centering}

\protect\caption{\emph{\footnotesize{}\label{tab:Tab Ib}Approximant of tails, case
}{\footnotesize{}$\beta\leq1$: Formulas used to adjust $\{\tau_{1,2},\lambda\}$
and their fitting constants. Corr. = correlation coefficient of nonlinear
fit. See left panel at figure \ref{Figure 05}.}}
\end{table}
\ref{tab:Tab Ib} describe a high frequency model and tables 
\begin{table}
\centering{}%
\begin{tabular}{cccc}
\hline 
\emph{\tiny{}Formulae} & {\tiny{}$\beta\exp\big[(\sum_{s=1}^{3}a_{s}(1-\beta)^{s})\exp(M\beta^{d})\big]$} &  & {\tiny{}$\frac{A}{\exp\big[(\sum_{s=1}^{3}a_{s}(1-\beta)^{s})\exp(M\beta^{d})\big]}$}\tabularnewline
\hline 
\hline 
{\footnotesize{}}%
\begin{tabular}{c}
{\footnotesize{}$\frac{Parameters}{Constants}$}\tabularnewline
{\footnotesize{}$M$}\tabularnewline
{\footnotesize{}$d$}\tabularnewline
{\footnotesize{}$a_{1}$}\tabularnewline
{\footnotesize{}$a_{2}$}\tabularnewline
{\footnotesize{}$a_{3}$}\tabularnewline
{\footnotesize{}Corr.}\tabularnewline
\end{tabular} & {\footnotesize{}}%
\begin{tabular}{c|c}
\multicolumn{1}{c}{{\footnotesize{}$\alpha_{1}$}} & {\footnotesize{}$\alpha_{2}$}\tabularnewline
{\footnotesize{}3.6903} & {\footnotesize{}2.42224}\tabularnewline
{\footnotesize{}1.4618} & {\footnotesize{}1.03791}\tabularnewline
{\footnotesize{}0.0190435} & {\footnotesize{}0.0962135}\tabularnewline
{\footnotesize{}0.0794805} & {\footnotesize{}0.193794}\tabularnewline
{\footnotesize{}0.572997} & {\footnotesize{}0.761107}\tabularnewline
\hline 
\multicolumn{1}{c}{\emph{\footnotesize{}0.999950}} & \emph{\footnotesize{}0.999942}\tabularnewline
\end{tabular} & {\footnotesize{}}%
\begin{tabular}{c}
{\footnotesize{}$\frac{Parameters}{Constants}$}\tabularnewline
{\footnotesize{}$M$}\tabularnewline
{\footnotesize{}$d$}\tabularnewline
{\footnotesize{}$A$}\tabularnewline
{\footnotesize{}$a_{1}$}\tabularnewline
{\footnotesize{}$a_{2}$}\tabularnewline
{\footnotesize{}$a_{3}$}\tabularnewline
{\footnotesize{}Corr.}\tabularnewline
\end{tabular} & {\footnotesize{}}%
\begin{tabular}{c|c}
\multicolumn{1}{c}{{\footnotesize{}$\gamma_{1}$}} & {\footnotesize{}$\gamma_{2}$}\tabularnewline
{\footnotesize{}4.54457} & {\footnotesize{}3.4752}\tabularnewline
{\footnotesize{}1.62482} & {\footnotesize{}1.47877}\tabularnewline
{\footnotesize{}0.999202} & {\footnotesize{}1.00287}\tabularnewline
{\footnotesize{}0.0227205} & {\footnotesize{}0.0401981}\tabularnewline
{\footnotesize{}-0.055399} & {\footnotesize{}0.054449}\tabularnewline
{\footnotesize{}0.79403} & {\footnotesize{}1.29633}\tabularnewline
\hline 
\multicolumn{1}{c}{\emph{\footnotesize{}0.998473}} & \emph{\footnotesize{}0.999654}\tabularnewline
\end{tabular}\tabularnewline
\hline 
\end{tabular}\protect\caption{\label{tab:Tab IIa}\emph{\footnotesize{}Approximant of heads, case
$\beta\leq1$}{\footnotesize{}: Formulas employed to adjust $\{\alpha_{1,2},\gamma_{1,2}\}$
and their fitting constants. Corr. = correlation coefficient. See
figure \ref{Figure 06}.}}
\end{table}
\ref{tab:Tab IIa} and \ref{tab:Tab IIb} do the same for a low frequency
one.
\begin{table}
\centering{}{\tiny{}}%
\begin{tabular}{ccc}
\hline 
{\tiny{}$\beta^{d}\sum_{s=0}^{4}b_{s}(1-\beta)^{s}$} & {\tiny{}$\sum_{s=1}^{2}\big\{ A_{s}(1-\beta)+B_{s}(1-\beta)^{2}\big\}\exp(-M_{s}\beta^{0.2})$} & {\tiny{}$\exp[-M(1-\beta)]\sum_{s=1}^{5}c_{s}(1-\beta)^{s}$}\tabularnewline
\hline 
\hline 
{\tiny{}}%
\begin{tabular}{c|c}
\multicolumn{1}{c}{{\footnotesize{}$\frac{Parameter}{Constants}$}} & {\footnotesize{}$\tau_{1}$}\tabularnewline
{\footnotesize{}$d$} & {\footnotesize{}0.785731}\tabularnewline
{\footnotesize{}$b_{0}$} & {\footnotesize{}0.726519}\tabularnewline
{\footnotesize{}$b_{1}$} & {\footnotesize{}0.160075}\tabularnewline
{\footnotesize{}$b_{2}$} & {\footnotesize{}7.16583}\tabularnewline
{\footnotesize{}$b_{3}$} & {\footnotesize{}-14.5815}\tabularnewline
{\footnotesize{}$b_{4}$} & {\footnotesize{}8.47342}\tabularnewline
\hline 
\multicolumn{1}{c}{{\footnotesize{}Corr.}} & \emph{\footnotesize{}0.997479}\tabularnewline
\end{tabular} & {\footnotesize{}}%
\begin{tabular}{c|c}
\multicolumn{1}{c}{{\footnotesize{}$\frac{Parameter}{Constants}$}} & {\footnotesize{}$\log_{10}\tau_{2}$}\tabularnewline
{\footnotesize{}$A_{1}$} & {\footnotesize{}42.4745}\tabularnewline
{\footnotesize{}$B_{1}$} & {\footnotesize{}-49.3268}\tabularnewline
{\footnotesize{}$M_{1}$} & {\footnotesize{}3.65441}\tabularnewline
{\footnotesize{}$A_{2}$} & {\footnotesize{}66182.6}\tabularnewline
{\footnotesize{}$B_{2}$} & {\footnotesize{}-61430.8}\tabularnewline
{\footnotesize{}$M_{2}$} & {\footnotesize{}12.7917}\tabularnewline
\hline 
\multicolumn{1}{c}{{\footnotesize{}Corr.}} & \emph{\footnotesize{}0.999987}\tabularnewline
\end{tabular} & {\footnotesize{}}%
\begin{tabular}{c|c}
\multicolumn{1}{c}{{\footnotesize{}$\frac{Parameter}{Constants}$}} & {\footnotesize{}$\lambda$}\tabularnewline
{\footnotesize{}$M$} & {\footnotesize{}4.60773}\tabularnewline
{\footnotesize{}$c_{1}$} & {\footnotesize{}4.80377}\tabularnewline
{\footnotesize{}$c_{2}$} & {\footnotesize{}-12.6617}\tabularnewline
{\footnotesize{}$c_{3}$} & {\footnotesize{}78.3249}\tabularnewline
{\footnotesize{}$c_{4}$} & {\footnotesize{}-120.494}\tabularnewline
{\footnotesize{}$c_{5}$} & {\footnotesize{}89.8475}\tabularnewline
\hline 
\multicolumn{1}{c}{{\footnotesize{}Corr.}} & \emph{\footnotesize{}0.996847}\tabularnewline
\end{tabular}\tabularnewline
\hline 
\end{tabular}\protect\caption{\emph{\footnotesize{}\label{tab:Tab IIb}Approximant of heads, case
$\beta\leq1$}{\footnotesize{}: Formulas employed to adjust $\{\tau_{1,2},\lambda\}$
and their fitting constants. Corr. = correlation coefficient. See
figure \ref{Figure 06}.}}
\end{table}

\subsection*{Case $2\ge\beta>1$}

\begin{table}
\centering{}\emph{\footnotesize{}}%
\begin{tabular}{cccc}
\hline 
\emph{\tiny{}Formulae} & {\tiny{}$1+\exp(\frac{-M}{\beta-1+\epsilon})\sum_{s=0}^{3}a_{s}(\beta-1)^{s},$} & {\tiny{}$\epsilon=10^{-180}.$} & {\tiny{}$A+\exp(M(\beta-1)^{3})\sum_{s=1}^{4}b_{s}(\beta-1)^{s}$}\tabularnewline
\hline 
\hline 
{\footnotesize{}}%
\begin{tabular}{c}
{\footnotesize{}$\frac{Parameters}{Constants}$}\tabularnewline
{\footnotesize{}$M$}\tabularnewline
{\footnotesize{}$a_{0}$}\tabularnewline
{\footnotesize{}$a_{1}$}\tabularnewline
{\footnotesize{}$a_{2}$}\tabularnewline
{\footnotesize{}$a_{3}$}\tabularnewline
{\footnotesize{}Corr.}\tabularnewline
\end{tabular} & {\footnotesize{}}%
\begin{tabular}{c|c}
\multicolumn{1}{c}{{\footnotesize{}$\alpha_{1}$}} & {\footnotesize{}$\alpha_{2}$}\tabularnewline
{\footnotesize{}0.116811} & {\footnotesize{}0.165105}\tabularnewline
{\footnotesize{}0.136577} & {\footnotesize{}-0.0645705}\tabularnewline
{\footnotesize{}-0.0269458} & {\footnotesize{}0.0697225}\tabularnewline
{\footnotesize{}-0.00692586} & {\footnotesize{}-0.0485374}\tabularnewline
{\footnotesize{}-0.00572689} & {\footnotesize{}0.0171122}\tabularnewline
\hline 
\multicolumn{1}{c}{\emph{\footnotesize{}0.999249}} & \emph{\footnotesize{}0.999926}\tabularnewline
\end{tabular} & {\footnotesize{}}%
\begin{tabular}{c}
{\footnotesize{}$\frac{Parameters}{Constants}$}\tabularnewline
{\footnotesize{}$A$}\tabularnewline
{\footnotesize{}$M$}\tabularnewline
{\footnotesize{}$b_{1}$}\tabularnewline
{\footnotesize{}$b_{2}$}\tabularnewline
{\footnotesize{}$b_{3}$}\tabularnewline
{\footnotesize{}$b_{4}$}\tabularnewline
{\footnotesize{}Corr.}\tabularnewline
\end{tabular} & {\footnotesize{}}%
\begin{tabular}{c|c}
\multicolumn{1}{c}{{\footnotesize{}$\gamma_{1}$}} & {\footnotesize{}$\gamma_{2}$}\tabularnewline
{\footnotesize{}1.90381} & {\footnotesize{}$\equiv$1.000}\tabularnewline
{\footnotesize{}2.25005} & {\footnotesize{}0.239754}\tabularnewline
{\footnotesize{}-4.09964} & {\footnotesize{}0.948757}\tabularnewline
{\footnotesize{}22.844} & {\footnotesize{}0.634155}\tabularnewline
{\footnotesize{}-29.9149} & {\footnotesize{}0.692054}\tabularnewline
{\footnotesize{}11.5881} & {\footnotesize{}-0.484575}\tabularnewline
\hline 
\multicolumn{1}{c}{\emph{\footnotesize{}0.999956}} & \emph{\footnotesize{}0.999993}\tabularnewline
\end{tabular}\tabularnewline
\hline 
\end{tabular}\protect\caption{\emph{\footnotesize{}\label{tab:Tab IIIa}Approximant for heads, case
$\beta>1$}{\footnotesize{}: Formulas used to adjust $\{\alpha_{1,2},\gamma_{1,2}\}$
and their fitting parameters. Corr. = correlation coefficient. The
point at $\beta=1$ is an outlier for a fit to $\gamma_{1}$ and is
removed. See figure \ref{Figure 07}.}}
\end{table}
Now the first pair of tables, 
\begin{table}
\centering{}{\tiny{}}%
\begin{tabular}{cc}
\hline 
{\tiny{}$c_{0}\sqrt{(\beta-1)}+\sum_{s=1}^{4}c_{s}(\beta-1)^{s}$} & {\tiny{}$\exp[-M(\beta-1)]\sum_{s=1}^{3}d_{s}(\beta-1)^{s}$}\tabularnewline
\hline 
\hline 
{\tiny{}}%
\begin{tabular}{cc|c}
{\footnotesize{}$\frac{Parameters}{Constants}$} & \multicolumn{1}{c}{{\footnotesize{}$-\log_{10}\tau_{1}$}} & {\footnotesize{}$-\log_{10}\tau_{2}$}\tabularnewline
{\footnotesize{}$c_{0}$} & {\footnotesize{}0.339472} & {\footnotesize{}0.27042}\tabularnewline
{\footnotesize{}$c_{1}$} & {\footnotesize{}0.27034} & {\footnotesize{}-0.1091}\tabularnewline
{\footnotesize{}$c_{2}$} & {\footnotesize{}0.0194173} & {\footnotesize{}0.977286}\tabularnewline
{\footnotesize{}$c_{3}$} & {\footnotesize{}0.26079} & {\footnotesize{}-0.884491}\tabularnewline
{\footnotesize{}$c_{4}$} & {\footnotesize{}-0.198826} & {\footnotesize{}0.303991}\tabularnewline
\cline{2-3} 
{\footnotesize{}Corr.} & \multicolumn{1}{c}{\emph{\footnotesize{}0.999853}} & \emph{\footnotesize{}0.999961}\tabularnewline
\end{tabular} & {\footnotesize{}}%
\begin{tabular}{c|c}
\multicolumn{1}{c}{{\footnotesize{}$\frac{Parameter}{Constants}$}} & {\footnotesize{}$\lambda$}\tabularnewline
{\footnotesize{}$M$} & {\footnotesize{}2.9177}\tabularnewline
{\footnotesize{}$d_{1}$} & {\footnotesize{}2.80567}\tabularnewline
{\footnotesize{}$d_{2}$} & {\footnotesize{}-3.18276}\tabularnewline
{\footnotesize{}$d_{3}$} & {\footnotesize{}4.6885}\tabularnewline
\hline 
\multicolumn{1}{c}{{\footnotesize{}Corr.}} & \emph{\footnotesize{}0.998497}\tabularnewline
\end{tabular}\tabularnewline
\hline 
\end{tabular}\protect\caption{\emph{\footnotesize{}\label{tab:Tab IIIb}Approximant for heads, case
$\beta>1$}{\footnotesize{}: Formulas used to adjust $\{\tau_{1,2},\lambda\}$
and their fitting parameters. Corr. = correlation coefficient. See
figure \ref{Figure 07}.}}
\end{table}
\ref{tab:Tab IIIa} and \ref{tab:Tab IIIb}, are those describing
a two-function approximant in the domain of low frequencies, and tables
\begin{table}
\begin{centering}
{\tiny{}}%
\begin{tabular}{cccc}
\hline 
\emph{\tiny{}Formulae} & {\tiny{}$\beta\big(1+a_{1}(\beta-1)\big)$} &  & {\tiny{}$\frac{A}{\big(1+a_{1}(\beta-1)\big)}$}\tabularnewline
\hline 
\hline 
{\footnotesize{}}%
\begin{tabular}{c}
{\footnotesize{}$\frac{Parameters}{Constants}$}\tabularnewline
{\footnotesize{}$a_{1}$}\tabularnewline
{\footnotesize{}Corr.}\tabularnewline
\end{tabular} & {\footnotesize{}}%
\begin{tabular}{cc}
{\footnotesize{}$\alpha_{1}$} & {\footnotesize{}$\alpha_{2}$}\tabularnewline
\multicolumn{1}{c|}{{\footnotesize{}-0.127772}} & {\footnotesize{}-0.00819886}\tabularnewline
\hline 
\emph{\footnotesize{}0.981962} & \emph{\footnotesize{}0.994925}\tabularnewline
\end{tabular} & {\footnotesize{}}%
\begin{tabular}{c}
{\footnotesize{}$\frac{Parameters}{Constants}$}\tabularnewline
{\footnotesize{}$A$}\tabularnewline
{\footnotesize{}$a_{1}$}\tabularnewline
{\footnotesize{}Corr.}\tabularnewline
\end{tabular} & {\footnotesize{}}%
\begin{tabular}{c|c}
\multicolumn{1}{c}{{\footnotesize{}$\gamma_{1}$}} & {\footnotesize{}$\gamma_{2}$}\tabularnewline
{\footnotesize{}$\equiv$2} & {\footnotesize{}$\equiv$1}\tabularnewline
{\footnotesize{}-0.110735} & {\footnotesize{}-0.0102285}\tabularnewline
\hline 
\multicolumn{1}{c}{\emph{\footnotesize{}0.833974}} & \emph{\footnotesize{}0.096606!}\tabularnewline
\end{tabular}\tabularnewline
\hline 
\end{tabular}
\par\end{centering}{\tiny \par}

\protect\caption{\emph{\footnotesize{}\label{tab:Tab IVa}Approximant for tails, case
$\beta>1$}{\footnotesize{}: Very rough trends decide formulas used
to adjust $\{\alpha_{1,2},\gamma_{1,2}\}$ since noise is greater
than signal at many points. The points at $\beta=$ 1.50, 1.52, 1.55,
1.80, 1.82 and 1.85 largely break the trend for a fit to $\alpha_{1}$
and $\gamma_{1}$ consequently they are removed. See figure \ref{Figure 08}.
Notice the strong linear feature of products $(\alpha\cdot\gamma)_{1,2}$
despite the noise (see figure \ref{Figure 09}).}}
\end{table}
\ref{tab:Tab IVa} and \ref{tab:Tab IVb} the corresponding to the
high frequency model.
\begin{table}
\centering{}\emph{\tiny{}}%
\begin{tabular}{cc}
\hline 
{\tiny{}$B+b_{0}\sqrt{\beta-1}+\sum_{s=1}^{4}b_{s}(\beta-1)^{s}$} & {\tiny{}$\exp[-M\sqrt{\beta-1}]\sum_{s=1}^{3}c_{s}(\beta-1)^{s}$}\tabularnewline
\hline 
\hline 
{\footnotesize{}}%
\begin{tabular}{cc|c}
{\footnotesize{}$\frac{Parameters}{Constants}$} & \multicolumn{1}{c}{{\footnotesize{}$-\log_{10}\tau_{1}$}} & {\footnotesize{}$-\log_{10}\tau_{2}$}\tabularnewline
{\footnotesize{}$A$} & {\footnotesize{}-0.278737} & {\footnotesize{}$\equiv$0}\tabularnewline
{\footnotesize{}$b_{0}$} & {\footnotesize{}1.69465} & {\footnotesize{}0.165075}\tabularnewline
{\footnotesize{}$b_{1}$} & {\footnotesize{}-4.114} & {\footnotesize{}0.219429}\tabularnewline
{\footnotesize{}$b_{2}$} & {\footnotesize{}9.33908} & {\footnotesize{}1.33322}\tabularnewline
{\footnotesize{}$b_{3}$} & {\footnotesize{}-10.9942} & {\footnotesize{}-2.39982}\tabularnewline
{\footnotesize{}$b_{4}$} & {\footnotesize{}4.63626} & {\footnotesize{}1.20034}\tabularnewline
\cline{2-3} 
{\footnotesize{}Corr.} & \multicolumn{1}{c}{\emph{\footnotesize{}0.996880}} & \emph{\footnotesize{}0.999016}\tabularnewline
\end{tabular} & {\footnotesize{}}%
\begin{tabular}{c|c}
\multicolumn{1}{c}{{\footnotesize{}$\frac{Parameter}{Constants}$}} & {\footnotesize{}$\lambda$}\tabularnewline
{\footnotesize{}$M$} & {\footnotesize{}-0.504694}\tabularnewline
{\footnotesize{}$c_{1}$} & {\footnotesize{}1.41581}\tabularnewline
{\footnotesize{}$c_{2}$} & {\footnotesize{}-1.49885}\tabularnewline
{\footnotesize{}$c_{3}$} & {\footnotesize{}0.572827}\tabularnewline
\hline 
\multicolumn{1}{c}{{\footnotesize{}Corr.}} & \emph{\footnotesize{}0.998585}\tabularnewline
\end{tabular}\tabularnewline
\hline 
\end{tabular}\protect\caption{\emph{\footnotesize{}\label{tab:Tab IVb}Approximant for tails, case
$\beta>1$}{\footnotesize{}: Formulas used to adjust $\{\tau_{1,2},\lambda\}$,
and their fitting parameters. The points at $\beta=$ 1.12, 1.15,
1.78, 1.80, 1.82 and 1.85 are not taking into account while adjusting.
See figure \ref{Figure 08}.}}
\end{table}

\newpage{}

\end{document}